\documentclass{svjour3}                     
\smartqed  
\usepackage{graphicx}
\usepackage{times}
\usepackage{subfigure}
\usepackage{paralist}
 \usepackage{amssymb}
 \usepackage{bbding}
 \usepackage{amsmath}
 \usepackage{appendix}
 \usepackage{booktabs}
 \usepackage{tabularx}
 \usepackage{multirow}
 \usepackage{bm}
\usepackage{latexsym}
\usepackage[colorlinks,
            linkcolor=blue,
            anchorcolor=green,
            citecolor=blue
            ]{hyperref}
\usepackage[sort&compress]{natbib}
\bibpunct{(}{)}{;}{a}{}{,}
 \journalname{Celestial Mechanics and Dynamical Astronomy}
\begin{document}

\title{Structures of secular resonances for inner test particles in hierarchical planetary systems 
}

\titlerunning{Structures of secular resonances}        


\author{Hanlun Lei}

\authorrunning{Lei} 


\institute{Hanlun Lei (\Envelope)\at
           School of Astronomy and Space Science, Nanjing University, Nanjing 210023, China\\
           Key Laboratory of Modern Astronomy and Astrophysics in Ministry of Education, Nanjing University, Nanjing 210023, China\\
           \email{leihl@nju.edu.cn}
}


\date{Received: date / Accepted: date}

\maketitle
\renewcommand{\arraystretch}{1.2}
\setlength{\arraycolsep}{5pt}
\begin{abstract}
In this study, dynamics of secular resonances for inner test particles are investigated under the octupole-level approximation by taking non-perturbative approaches. In practice, webs of the major secular resonances are produced by identifying families of stable periodic orbits and the associated stable libration zones are obtained by analysing Poincar\'e surfaces of section. By taking different values of the factor $\epsilon$ ($\epsilon$ measures the contribution of octupole terms), the influences of the octupole-order terms upon the dynamical structures are evaluated. Under the condition of $\epsilon = 0$ (no octupole-order contribution), the dynamical model is totally integrable and there is only Kozai resonance arising in the phase space. When the factor $\epsilon$ is different from zero, the dynamical structure in the phase space becomes complicated due to varieties of secular resonances appearing. Numerical results further indicate that (a) distributions of libration centres and stable libration zones remain qualitatively similar with different values of $\epsilon$, (b) Kozai resonance disappears due to the chaotic motion in the low-eccentricity region, and (c) the chaotic area arising in the low-eccentricity region increases with the factor $\epsilon$. Secular resonances are the source of many important dynamical phenomena, such as chaos, orbit alignment and orbit flipping, and thus the results presented in this work could be useful to understand the secular dynamics for those high-eccentricity and/or high-inclination objects in hierarchical planetary systems.

\end{abstract}

\section{Introduction}
\label{Sect1}

Hierarchical planetary systems are common in the Universe, covering a large range of astrophysical systems from the satellite and planet scales to supermassive black hole \citep{naoz2016eccentric}. In particular, when one of the triple system is assumed as a massless particle, the system reduces to the so-called hierarchical restricted three-body problem.

To explore the secular dynamics of test particles on timescales much longer than their orbital periods, the short-period effects over the orbital periods of the inner and outer binaries are averaged out by means of averaging approach \citep{lidov1962evolution, kozai1962secular, ford2000secular, naoz2013secular, luo2016double, lei2019semi, lei2020dynamical}. Under the standard double-averaged model, \citet{lidov1962evolution} studied the secular dynamics of artificial satellites perturbed by the Sun and Moon and, at the same age,  \citet{kozai1962secular} investigated the secular influences of Jupiter's gravitational perturbations upon those highly inclined and eccentric asteroids in our Solar system. It is known that, when the inclination is greater than $39.2^{\circ}$, the eccentricity and inclination of satellites or asteroids would undergo a coupled oscillation \citep{lidov1962evolution, kozai1962secular, thomas1996kozai}. The coupled oscillation between eccentricity and inclination is due to the existence of Kozai resonance, which happens between the longitude of pericentre $\varpi$ and longitude of ascending node $\Omega$. The resulting mechanism exciting eccentricity and/or inclination of test particles is called von Zeipel--Lidov--Kozai mechanism \citep{ito2019lidov}. In both \citet{lidov1962evolution} and \citet{kozai1962secular}, the following two assumptions are made: (a) the test-particle approximation for the satellites or asteroids and (b) circular-orbit approximation for the orbit of perturber. The circular-orbit approximation would lead to axisymmetric potential where the longitude of ascending node $\Omega$ disappears from the Hamiltonian, so that the dynamical model becomes an integrable system. Usually, the double-averaged disturbing function is truncated at the lowest order in the semimajor axis ratio, which is called the test particle quadrupole approximation (TPQ) by \citet{naoz2016eccentric}.

However, if the circular-orbit approximation is relaxed, the next level (octupole order) of approximation needs to be taken into consideration and the resulting dynamical model is of two degrees of freedom with $\Omega$ and $\omega$ as angular coordinates. Under such a relaxation, the standard von Zeipel--Lidov--Kozai mechanism naturally becomes the eccentric version.

According to the octupole-level secular perturbation theory, \citet{lee2003secular} studied the apsidal secular resonance with critical argument of $\sigma = \varpi_1 - \varpi_2$ under coplanar, hierarchical, two-planet systems and they showed that this resonance causes large amplitude variations of eccentricities. Under the octupole-order approximation, \citet{lithwick2011eccentric} proved that the particle's vertical angular momentum is no longer a constant and its Kozai oscillations are modulated on timescales much longer than Kozai period. Particularly, the modulation of Kozai cycles could make the particle's orbit flip from prograde to retrograde and back again, and at the flipping instant the orbit could reach arbitrarily high eccentricities (this is called the eccentric von Zeipel--Lidov--Kozai mechanism). Such an eccentric von Zeipel--Lidov--Kozai mechanism provides a possible clue for the formation of retrograde `hot Jupiters' \citep{naoz2011hot}: (a) the secular perturbation coming from the perturber on an elliptic orbit causes orbit flipping, which is accompanied by an excursion to high eccentricity and thus low altitude of periastron, and (b) strong planet--star tidal interactions damp the eccentricity of the inner orbit, forming a retrograde hot Jupiter. Considering the fact that the octupole-order potential causes a slow and cyclic modulation of the Kozai--Lidov cycles, \citet{katz2011long} studied the very long-term evolution analytically by averaging the secular equations of motion over the Kozai--Lidov cycles (at the same time introducing a new constant of motion) and derived an analytical criterion for orbit flipping. \citet{li2014chaos} studied the eccentric von Zeipel--Lidov--Kozai mechanism systematically by taking advantage of Poincar\'e surfaces of section and the Lyapunov exponents and they found that, when the mutual inclination is high, the resonances due to the octupole potential are the main causes that lead to orbits flipping, significant eccentricity excitation, and chaotic behaviors. \citet{sidorenko2018eccentric} interpreted the eccentric von Zeipel--Lidov--Kozai effect that causes orbit flipping under the octupole-order approximation as a resonance phenomenon with the critical argument $\sigma = \Omega + {\rm sign}(\cos{i}) \omega$ librating around $0^{\circ}$ or $180^{\circ}$ ($\rm sign$ is the sign function, $\Omega$ the longitude of ascending node, $\omega$ the argument of pericentre and $i$ the inclination).

Regarding the dynamical model truncated at the octupole order, the previous studies indicate that the secular resonances govern the long-term dynamics, including the coupled excitation of eccentricity and inclination, orbit flipping and chaotic behaviours. However, no systematic mapping of the secular resonances has been conducted yet. In this context, we explore the global dynamical structures (or webs of secular resonances) in the phase space with an aim at understanding the secular dynamics for those high-eccentricity and/or high-inclination objects in hierarchical restricted three-body systems.

The remaining part of this work is organised as follows. In Sect. \ref{Sect2}, the dynamical model for long-term evolutions of test particles is briefly introduced and Kozai dynamics under the quadrupole-order approximation is discussed and, in Sect. \ref{Sect3}, the Poincar\'e surfaces of sections at different levels of Hamiltonian are produced and analysed. In Sect. \ref{Sect4}, families of symmetric periodic orbits (or webs of secular resonances) are determined and, in Sect. \ref{Sect5}, the boundaries of stable libration zones are produced with different system parameters. Finally, the conclusion is summarised in Sect. \ref{Sect6}.

\section{Hamiltonian model and Kozai dynamics}
\label{Sect2}

The dynamical model truncated at the octupole order is briefly presented, and then under the quadrupole-level approximation the Kozai dynamics is recalled for the purpose of comparison.

\subsection{Hamiltonian function}
\label{Sect2_1}

We take the hierarchical restricted three-body problem composed of two primaries (a central body and a distant planet) and an inner test particle as the fundamental model. Under this model, the primaries move around their barycentre in fixed elliptic orbits and the test particle moves around the central body in a perturbed Keplerian orbit.

To describe the orbits of test particles and perturber, we choose an inertial right-handed reference frame originated at the central body, denoted by $o$-$xyz$, with the perturber's orbit as the fundamental plane, the $z$-axis along the total angular momentum vector and the $x$-axis along the eccentricity vector of the perturber. Under the coordinate system, the orbits of the test particle (and the perturber) are characterised by six orbital elements, including the semimajor axis $a$ ($a_p$), eccentricity $e$ ($e_p$), inclination $i$ ($i_p$), longitude of ascending node $\Omega$ ($\Omega_p$), argument of pericentre $\omega$ ($\omega_p$) and the mean anomaly $M$ ($M_p$). Unless otherwise specified, in the entire work we utilise the variables with subscript `$p$' for the perturber and the ones without any subscript for test particles. Due to the choice of the reference plane and the $x$-axis, both the inclination of perturber $i_p$ and the longitude of pericentre $\varpi_p$ are zero.

In hierarchical configurations, it requires that the semimajor axis of the test particle is much smaller than that of the secondary, i.e., the semimajor axis ratio $\alpha = a/a_p$ is a small parameter. Following \citet{harrington1968dynamical, harrington1969stellar}, the Hamiltonian, describing the motion of test particles perturbed by a distant body, can be written as
\begin{equation}\label{Eq1}
{\cal H} =  - \frac{\mu }{{2a}} - \frac{{{\cal G}{m_p}}}{{{a_p}}}\sum\limits_{n = 2}^\infty  {{\alpha ^n}{{\left( {\frac{r}{a}} \right)}^n}{{\left( {\frac{{{a_p}}}{{{r_p}}}} \right)}^{n + 1}}{P_n}\left( {\cos \psi } \right)},
\end{equation}
where $\mu = {\cal G} m_0$ is the gravitational parameter of the central body, $m_0$ and $m_p$ are the mass of the central body and perturber, $r$ and $r_p$ are the distances of the test particle and perturber relative to the central body, $\psi$ stands for the angular separation between their position vectors, and $P_n(\cos{\psi})$ are the Legendre polynomial of $\cos{\psi}$ with degree $n$. The second term in the right hand of equation (\ref{Eq1}) is called the disturbing function.

To study the secular dynamics of test particles in hierarchical systems, it is required to make a secular approximation for the Hamiltonian function \citep{ford2000secular,naoz2013secular,naoz2016eccentric}. Usually, secular approximation is achieved by means of phase averaging over the orbital periods of the inner and outer binaries as follows:
\begin{equation*}
{\cal H}^*  = \frac{1}{{4{\pi ^2}}}\int\limits_0^{2\pi } {\int\limits_0^{2\pi } {{\cal H}{\rm d}M{\rm d}{M_p}} }.
\end{equation*}
To make sure that the double-averaged Hamiltonian could effectively predict the long-term evolution of the test particle, it is required that the mass of the perturber should be much smaller than that of the central body. Otherwise, the effects filtered by double averaging need to be taken into account in formulating the dynamical model of long-term evolution \citep{cuk2004secular, luo2016double, lei2018modified, lei2019semi, hamers2019analytic}.

In the averaged model, the mean anomaly of test particle disappears from the Hamiltonian, so that its semimajor axis remains unchanged in the long-term evolution. Let us further truncate the Hamiltonian at the third order in $\alpha$ and eliminate those constant terms, so that the octupole-level approximation of secular Hamiltonian becomes \citep{ford2000secular, lithwick2011eccentric, naoz2016eccentric}
\begin{equation*}
{\cal H}^* =  - {{\cal C}_0}\left( {{F_{\rm quad}} + \epsilon {F_{\rm oct}}} \right)
\end{equation*}
where the coefficient ${{\cal C}_0}$ is
\begin{equation*}
{{\cal C}_0} = \frac{3}{8}\frac{{{\cal G}{m_p}}}{{{a_p}}}{\left( {\frac{a}{{{a_p}}}} \right)^2}\frac{1}{{{{\left( {1 - e_p^2} \right)}^{3/2}}}},
\end{equation*}
the quadrupole- and octupole-order terms ${F_{\rm quad}}$ and ${F_{\rm oct}}$ are given by
\begin{equation*}
{F_{\rm quad}} =  - \frac{1}{2}{e^2} + {\cos ^2}i + \frac{3}{2}{e^2}{\cos ^2}i + \frac{5}{2}{e^2}\left( {1 - {{\cos }^2}i} \right)\cos \left( {2\omega } \right),
\end{equation*}
\begin{equation*}
\begin{aligned}
{F_{\rm oct}} & =  \frac{5}{{16}}\left( {e + \frac{3}{4}{e^3}} \right) \left( {1 - 11\cos i - 5{{\cos }^2}i+15{{\cos }^3}i} \right)\cos \left( {\omega  - \Omega } \right)\\
&+\frac{5}{{16}}\left( {e + \frac{3}{4}{e^3}} \right) \left( {1 + 11\cos i - 5{{\cos }^2}i - 15{{\cos }^3}i} \right)\cos \left( {\omega  + \Omega } \right)\\
& - \frac{{175}}{{64}}{e^3} \left( {1 - \cos i - {{\cos }^2}i + {{\cos }^3}i} \right)\cos \left( {3\omega  - \Omega } \right)\\
& - \frac{{175}}{{64}}{e^3} \left( {1 + \cos i - {{\cos }^2}i - {{\cos }^3}i} \right)\cos \left( {3\omega  + \Omega } \right)
\end{aligned}
\end{equation*}
and the factor $\epsilon$, measuring the significance of octupole-order term, is given by \citep{lithwick2011eccentric}
\begin{equation*}
\epsilon  = \frac{a}{{{a_p}}}\frac{{{e_p}}}{{1 - e_p^2}} = \alpha \frac{{{e_p}}}{{1 - e_p^2}}.
\end{equation*}
For the same dynamical model, the double-averaged Hamiltonian can be formulated up to an arbitrary order in the semimajor axis ratio $\alpha$, as given by \citet{lei2021}.

\begin{figure*}
\centering
\includegraphics[width=0.8\textwidth]{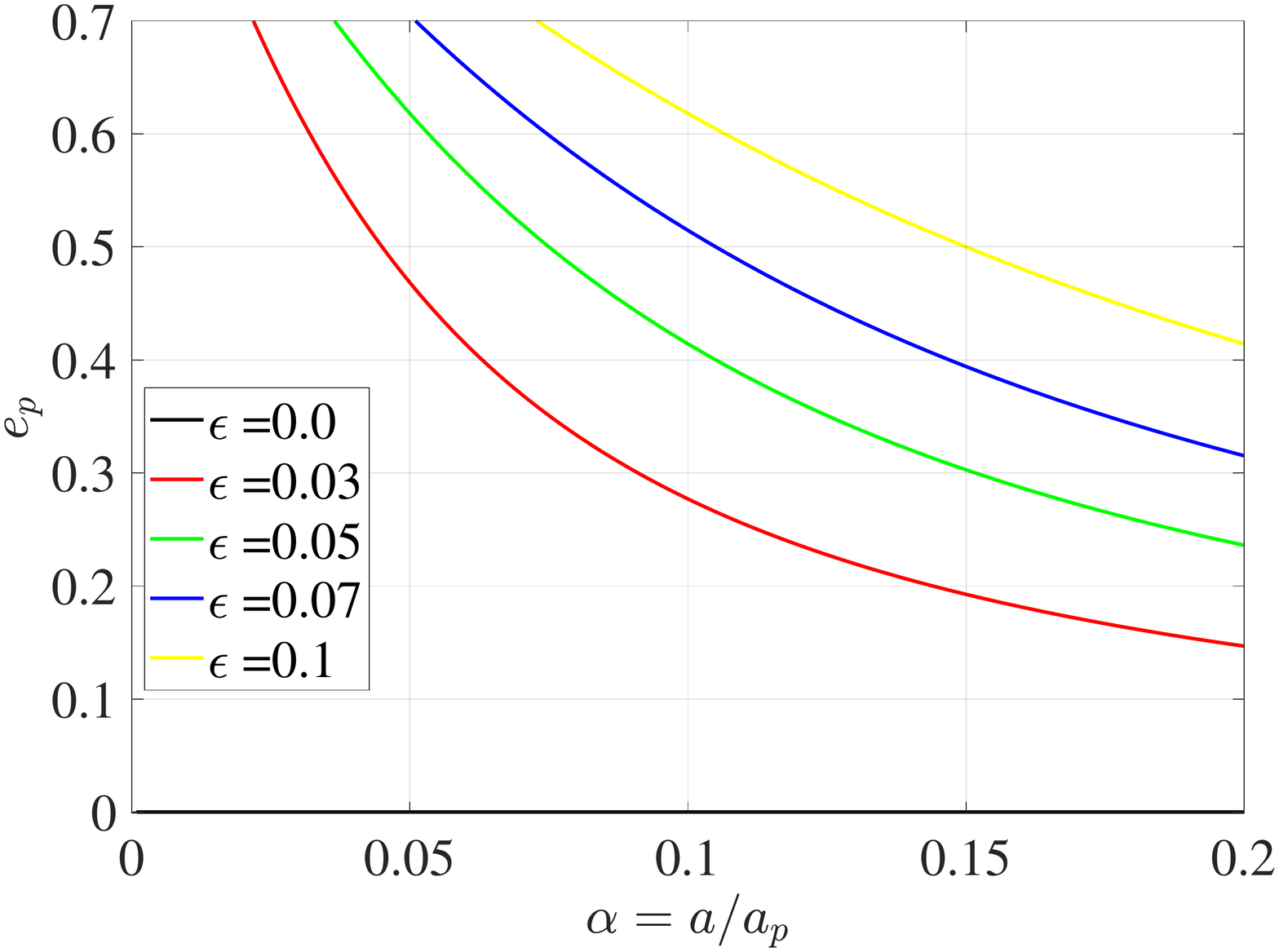}
\caption{Representative values of the system parameter $\epsilon$ adopted in this study. As a special case, the dynamical model specified by $\epsilon = 0$ will be used in Sect. \ref{Sect2_2}. Dynamical models specified by $\epsilon = 0.03, 0.05, 0.07, 0.1$ are to be used in Sects \ref{Sect3}--\ref{Sect5}.}
\label{Fig8}
\end{figure*}

Evidently, the octupole-level contribution to the secular Hamiltonian is determined by the magnitude of $\epsilon$, which is positively correlated to the semimajor axis ratio $\alpha = a/a_p$ and the eccentricity of the perturber's orbit $e_p$. In Fig. \ref{Fig8}, several level curves of $\epsilon$ are given in the space $(\alpha, e_p)$. Obviously, each value of $\epsilon$ stands for a series of dynamical models. In particular, when the perturber is on a circular orbit (corresponding to $\epsilon = 0$), the octupole-level terms vanish. In this case, the dynamical model reduces to the well-known circular restricted three-body problem (CRTBP), in which the octupole-order term has no contribution.

Considering the fact that the coefficient ${\cal C}_0$ is a constant in the long-term evolution, we can scale the Hamiltonian ${\cal H}^*$ by ${\cal C}_0$ as follows (for simplicity, the notation ${\cal H}$ is used to represent the scaled Hamiltonian):
\begin{equation}\label{Eq2}
{\cal H} = {{\cal H}^*} / {{\cal C}_0} =  - \left( {{F_{\rm quad}} + \epsilon {F_{\rm oct}}} \right).
\end{equation}
The normalised Hamiltonian given by equation (\ref{Eq2}) has been widely used in previous works \citep{lithwick2011eccentric, li2014chaos, naoz2016eccentric}. The scaling of the Hamiltonian given by Eq. (\ref{Eq2}) is equivalent to changing the unit of time.

To formulate the equations of motion, let's adopt the scaled Delaunay variables (the action variables are scaled by the constant factor $\sqrt{\mu a}$) in the following manner \citep{lithwick2011eccentric}:
\begin{equation*}
\begin{aligned}
&g = \omega,\quad G = \sqrt {1 - {e^2}},\\
&h = \Omega,\quad H = G\cos i.
\end{aligned}
\end{equation*}
Thus, the Hamiltonian ${\cal H}$ can be re-organised as a function of the defined Delaunay variables by ${\cal H} (G,H,g,h)$. The dynamical model holds the following symmetric properties:
\begin{equation*}
{\cal H} (G,H,g,h) = {\cal H} (G,-H,2\pi - g,h) = {\cal H} (G,-H,g,2\pi - h)= {\cal H} (G,H,g + \pi,h + \pi)
\end{equation*}
where the first two symmetries can be found in \citet{sidorenko2018eccentric}. The (scaled) Hamiltonian canonical relations hold
\begin{equation}\label{Eq3}
\begin{aligned}
\frac{{{\rm d}g}}{{{\rm d}t}} =& \frac{{\partial {\cal H}}}{{\partial G}},\quad \frac{{{\rm d}G}}{{{\rm d}t}} =  - \frac{{\partial {\cal H}}}{{\partial g}},\\
\frac{{{\rm d}h}}{{{\rm d}t}} =& \frac{{\partial {\cal H}}}{{\partial H}},\quad \frac{{{\rm d}H}}{{{\rm d}t}} =  - \frac{{\partial {\cal H}}}{{\partial h}},
\end{aligned}
\end{equation}
where the scaled time is \citep{lithwick2011eccentric}
\begin{equation*}
\begin{aligned}
t &= \frac{{\cal C}_0}{\sqrt{\mu a}}  t_r\\
& = \frac{3}{8}\frac{{{m_p}}}{{{m_0}}}n{\left( {\frac{a}{{{a_p}}}} \right)^3}\frac{1}{{{{\left( {1 - e_p^2} \right)}^{3/2}}}}  {t_r}
\end{aligned}
\end{equation*}
with $n$ as the mean motion frequency of the test particle moving around the central body and $t_r$ as the true time.

To remove the singularity of variables at small eccentricities and inclinations, the following variables are also used for describing the motion of test particles:
\begin{equation*}
\begin{aligned}
k &= e \cos \omega,\quad h = e \sin \omega,\\
q &= \sin{\frac{i}{2}} \cos \Omega,\quad p = \sin{\frac{i}{2}} \sin \Omega.
\end{aligned}
\end{equation*}
It is not difficult to realise the transformations among the following sets of variables:
\begin{equation*}
(e,i,\Omega,\omega)  \Leftrightarrow (G,H,g,h) \Leftrightarrow (k,h,q,p).
\end{equation*}

\subsection{Kozai dynamics}
\label{Sect2_2}

For completeness, here we make some discussions about Kozai dynamics in the special case of $\epsilon = 0$, which is equivalent to the quadrupole-level approximation. When $\epsilon = 0$, the (scaled) Hamiltonian given by Eq. (\ref{Eq2}) can be organised as
\begin{equation}\label{Eq4}
{\cal H} = \frac{1}{2}\left( {1 - {G^2} + 3{H^2} - 5\frac{{{H^2}}}{{{G^2}}}} \right) - \frac{5}{2}\left( {1 - {G^2} + {H^2} - \frac{{{H^2}}}{{{G^2}}}} \right)\cos \left( {2g} \right).
\end{equation}
The angular variable $h$ is a cyclic variable, showing that its conjugate momentum $H = \sqrt{1-e^2}\cos{i}$ becomes a motion integral under the quadrupole-order approximation. Naturally, the dynamical model represented by Eq. (\ref{Eq4}) is of one degree of freedom. Similar to \citet{kozai1962secular}, let us denote the motion integral by
\begin{equation}\label{Eq5}
H = \sqrt{1-e^2}\cos{i} = \cos{i_*},
\end{equation}
where $i_*$ stands for the critical inclination (known as Kozai parameter) when the eccentricity is assumed at zero.

The equilibrium points should satisfy the following stationary condition:
\begin{equation}\label{Eq6}
\begin{aligned}
\dot g &=  - G + 5\frac{{{H^2}}}{{{G^3}}} + 5\left( {G - \frac{{{H^2}}}{{{G^3}}}} \right)\cos \left( {2g} \right) = 0,\\
\dot G &=  - 5\left( {1 - {G^2} + {H^2} - \frac{{{H^2}}}{{{G^2}}}} \right)\sin \left( {2g} \right) = 0.
\end{aligned}
\end{equation}
The second equality of Eq. (\ref{Eq6}) indicates that the equilibrium points are located at $2g = 0$ or $2g = \pi$. It is not difficult to demonstrate that the equilibrium points at $2g = \pi$ are stable in dynamics (because the Hamiltonian takes its maximum value when $2g=\pi$), showing that Kozai resonance occurs at $2g=\pi$.

At the Kozai centre ($2g=\pi$), the first equality of Eq. (\ref{Eq6}) yields
\begin{equation}\label{Eq8}
G^4 = \frac{5}{3} H^2,
\end{equation}
which leads to the following relation between eccentricity and inclination at Kozai centres \citep{kozai1962secular},
\begin{equation}\label{Eq9}
\cos{i} = \pm \sqrt{\frac{3}{5}(1-e^2)}.
\end{equation}
Eq. (\ref{Eq9}) determines two curves of Kozai centre in the eccentricity--inclination space, one is in the prograde region and the other one is in the retrograde region (see Fig. \ref{Fig7_1}). Furthermore, the eccentricity of Kozai centre can be expressed as a function of the Kozai parameter $i_*$,
\begin{equation}\label{Eq11}
e = \sqrt{1-\sqrt{\frac{5}{3}\cos^2{i_*}}},
\end{equation}
and the inclination of Kozai centre is expressed as a function of $i_*$,
\begin{equation}\label{Eq12}
\quad \cos{i} = \sqrt{\frac{3}{5}} \cos{i_*}.
\end{equation}
Physical solution of Eq. (\ref{Eq11}) implies that the Kozai parameter should satisfy $\cos^2{i_*} < 3/5$ (i.e., $\left|H\right| < \sqrt{3/5}$ or $39.23^{\circ} < i_* < 140.77^{\circ}$). Thus, the occurrence of Kozai resonance requires $39.23^{\circ} < i_* < 140.77^{\circ}$ (or $\left|H\right| < \sqrt{3/5}$), meaning that Kozai resonance disappears when $i_*$ is smaller than $39.23^{\circ}$ or greater than $140.77^{\circ}$.

\begin{figure*}
\centering
\includegraphics[width=0.49\textwidth]{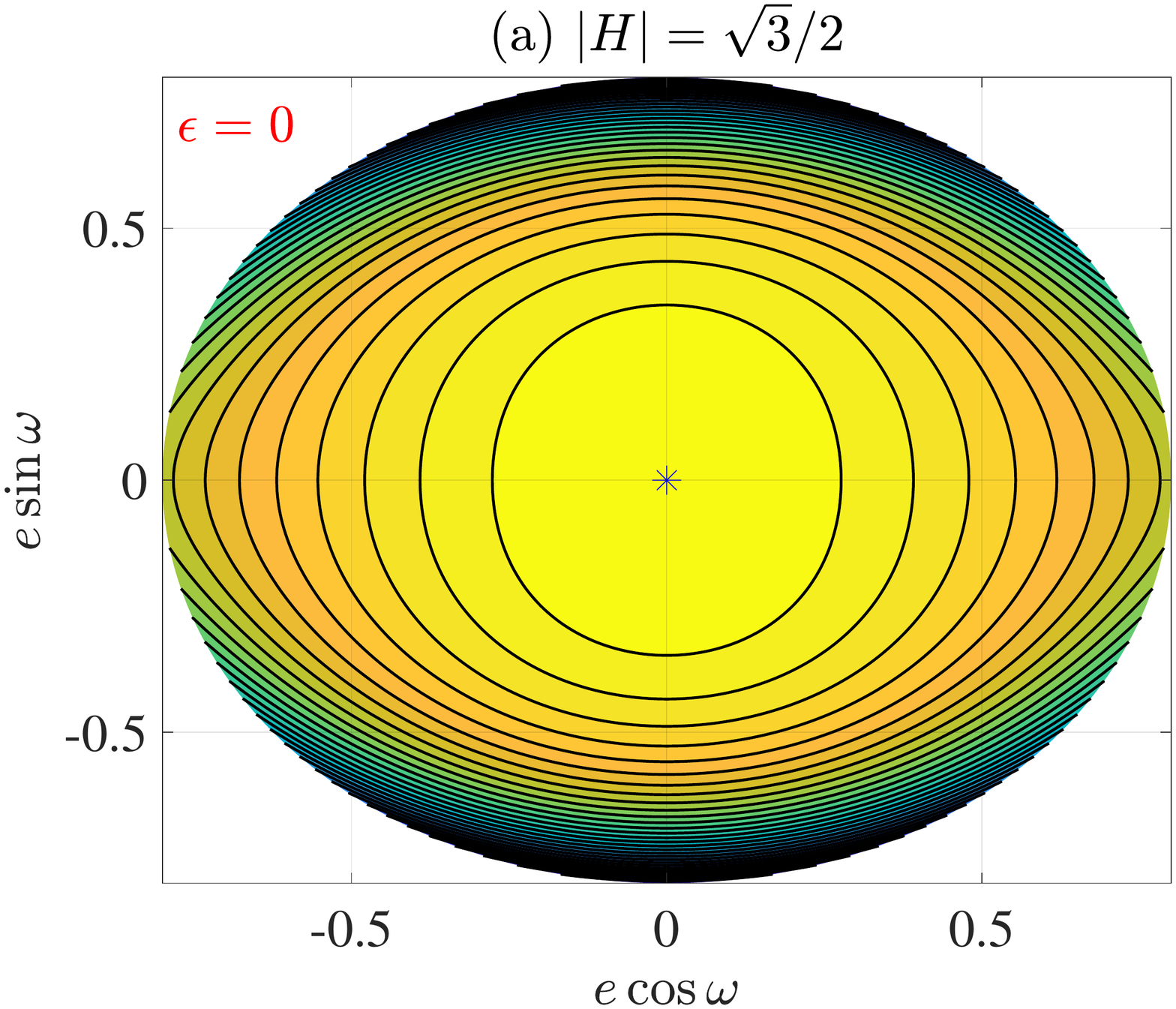}
\includegraphics[width=0.49\textwidth]{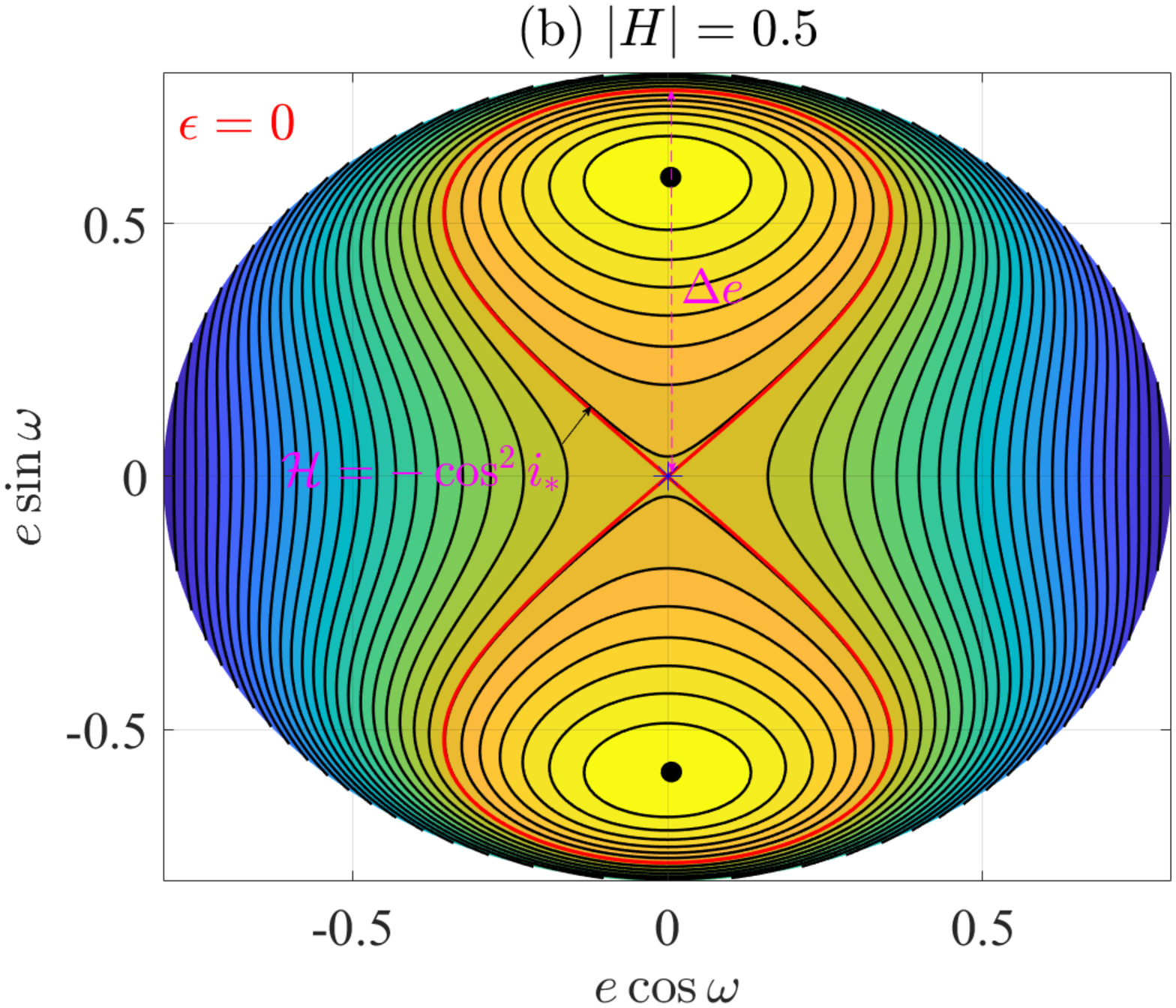}
\caption{Phase portraits of the dynamics at the quadrupole order specified by the motion integral $\left|H\right| = \sqrt{3}/2$ corresponding to the Kozai parameter $i_* = 30^{\circ}$ or $i_* = 150^{\circ}$ (\emph{left panel}) and $\left|H\right| = 0.5$ corresponding to the Kozai parameter $i_* = 60^{\circ}$ or $i_* = 120^{\circ}$ (\emph{right panel}). The zero-eccentricity point (i.e., the coordinate origin) is marked by a black star. In the \emph{left panel}, all motions are of circulation. In the \emph{right panel}, there are three equilibrium points and the regions of libration and circulation are divided by the dynamical separatrix (shown by the red line). The separatrix corresponds to the level curve of ${\cal H} = -H^2 = -\cos^2{i_*}$.}
\label{Fig6}
\end{figure*}

In Fig. \ref{Fig6}, we report the phase portraits (i.e., level curves of Hamiltonian) in the space $(e\cos{\omega},e\sin{\omega})$ for the motion integral at $\left|H\right| = \sqrt{3}/2$ (corresponding to the Kozai parameter $i_* = 30^{\circ}$ or $i_* = 150^{\circ}$) and at $\left|H\right| = 0.5$ (corresponding to the Kozai parameter $i_* = 60^{\circ}$ or $i_* = 120^{\circ}$). In the case of $\left|H\right| = \sqrt{3}/2$ (see the left panel of Fig. \ref{Fig6}), Kozai resonance cannot happen and all the motions in the phase space are of circulation. In the case of $\left|H\right| = 0.5$ (see the right panel of Fig. \ref{Fig6}), the Kozai resonance takes place around $2\omega = \pi$ and the regions of libration and circulation are divided by the dynamical separatrix.

For the one-degree-of-freedom dynamical model at hand, the Hamiltonian ${\cal H}$ and the motion integral $H$ (or the Kozai parameter $i_*$) are two conserved quantities and their magnitudes determine the mode of motion (libration or circulation). Next, the libration zones of Kozai resonance are discussed.

On one hand, let us discuss the circulation and libration zones of Kozai resonance in the space spanned by ${\cal H}$ and $H$ (or $i_*$). According to the Hamiltonian given by Eq. (\ref{Eq4}), we can get that the minimum Hamiltonian happens at $G = \left|H\right|$ (i.e., the inclination $i$ is equal to $0^{\circ}$ or $180^{\circ}$) and $2g = 0$,
\begin{equation}\label{Eq14}
{\cal H} = -2 + H^2 = -2 + \cos^2{i_*},
\end{equation}
which provides a lower boundary of the allowed Hamiltonian ${\cal H}$ as a function of $H$ (or $i_*$). According to resonant theory, the Hamiltonian takes its maximum value at the resonant centre (stable equilibrium point). Accordingly, we get that the maximum Hamiltonian happens at the Kozai centre (i.e., $2g = \pi$ and $G^4 = 5/3 H^2$),
\begin{equation}\label{Eq15}
{\cal H} = 3 + 4 H^2 - 2\sqrt{15}\left|H\right| = 3 + 4 \cos^2{i_*} - 2\sqrt{15}\left|\cos{i_*}\right|,
\end{equation}
which provides the upper boundary of the allowed Hamiltonian ${\cal H}$ as a function of $H$ (or $i_*$). Additionally, it is observed from the phase portrait (see the right panel of Fig. \ref{Fig6}) that the separatrix passes through the zero-eccentricity point. Thus, the dynamical separatrix holds the Hamiltonian
\begin{equation}\label{Eq16}
{\cal H} = -H^2 = -\cos^2{i_*},
\end{equation}
which provides the boundary between the libration and circulation zones in the space $({\cal H}, H)$ or $({\cal H}, i_*)$.

\begin{figure*}
\centering
\includegraphics[width=0.8\textwidth]{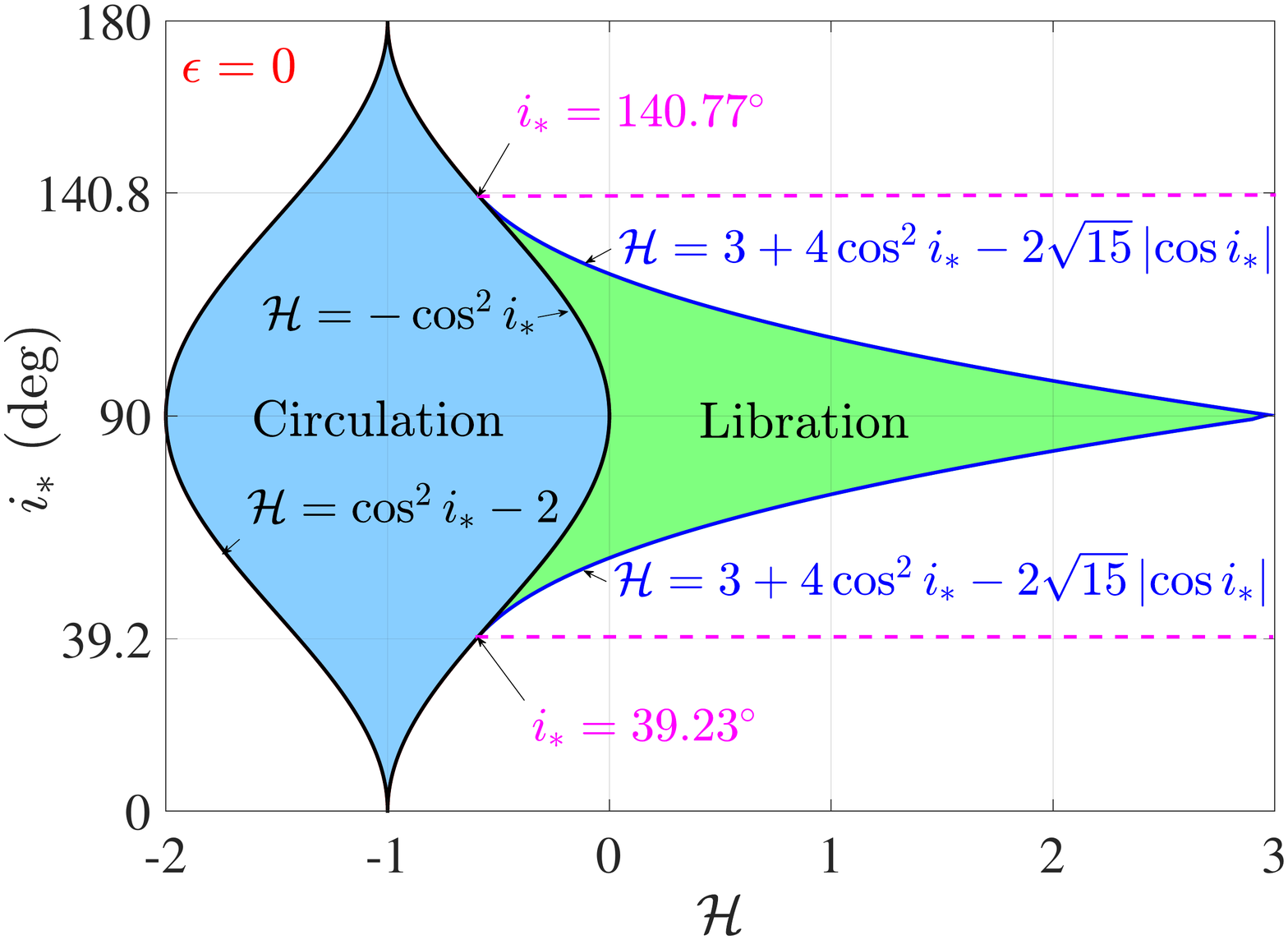}
\caption{Distribution of libration and circulation domains in the space spanned by Hamiltonian ${\cal H}$ and Kozai parameter $i_*$. The critical lines of $i_* = 39.23^{\circ}, 140.77^{\circ}$ are marked.}
\label{Fig7}
\end{figure*}

In Fig. \ref{Fig7}, the lower and upper boundaries of the allowed regions and the separatrix dividing libration and circulation zones are reported in the space $({\cal H}, i_*)$. The allowed regions are bounded by the lines corresponding to ${\cal H} = \cos^2{i_*} - 2$ and ${\cal H} = 3 + 4\cos^2{i_*}-2\sqrt{15}\left|\cos{i_*}\right|$ and the white areas in the space are forbidden regions. The circulation and libration zones are separated by the critical line of ${\cal H} = -\cos^2{i_*}$ (called separatrix). From Fig. \ref{Fig7}, it is observed that the occurrence of Kozai resonance requires that the Kozai parameter and Hamiltonian should satisfy the following conditions: $39.23^{\circ} < i_* < 140.77^{\circ}$ and $-0.6 < {\cal H} < 3$.

About the distribution of libration and circulation zones, similar plot shown in the space $({\cal H}, \left|H\right|)$ has been obtained in \citet{sidorenko2018eccentric}, who derived the boundaries by introducing auxiliary functions.

\begin{figure*}
\centering
\includegraphics[width=0.8\textwidth]{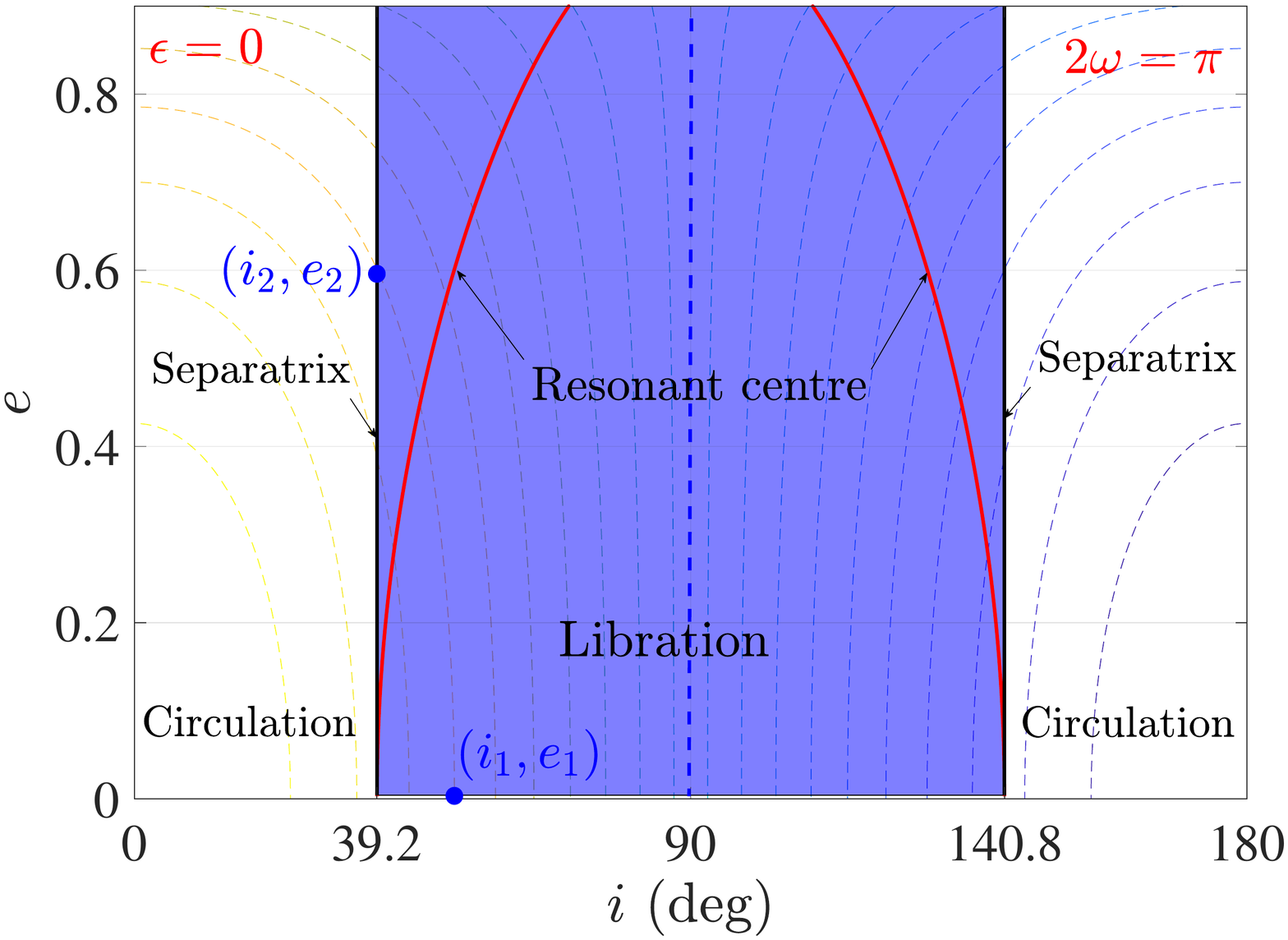}
\caption{Distribution of libration and circulation domains in the eccentricity--inclination space. The level curves of the motion integral $H$ are plotted and the polar line is marked by a vertical blue dashed line. The separatrices corresponding to $i = 39.23^{\circ}$ and $i = 140.77^{\circ}$ are shown by two vertical black lines.}
\label{Fig7_1}
\end{figure*}

On the other hand, let us discuss the distribution of libration and circulation zones of Kozai resonance in the eccentricity--inclination space. As shown in the right panel of Fig. \ref{Fig6}, the width of Kozai resonance can be measured at $2\omega = 180^{\circ}$ (location of resonant centre) and it is denoted by the variation of eccentricity $\Delta e = e_{2} - e_{1}$ (the resonant width in terms of inclination variation can be naturally denoted by $\Delta i = i_2 - i_1$).

Along a certain isoline specified by $H$ or $i_*$, we can see that the lower boundary of eccentricity reaches $e_{1} = 0$ where the inclination is equal to the Kozai parameter $i_1 = i_*$ (see Fig. \ref{Fig7_1}). About the eccentricity $e_2$ at the upper boundary (the action variable is denoted by $G_2 = \sqrt{1-e_2^2}$), it satisfies the following equality (see the phase portrait shown in the right panel of Fig. \ref{Fig6}),
\begin{equation*}
{\cal H} \left(H=\cos{i_*}; 2g=\pi, G=G_2\right) = {\cal H} \left(H=\cos{i_*}; 2g=0, G=1\right),
\end{equation*}
which leads to the eccentricity at the upper boundary \citep{naoz2016eccentric},
\begin{equation}\label{Eq14}
e_2 = \sqrt{1-\frac{5}{3}\cos^2{i_*}}.
\end{equation}
The motion integral $H = \sqrt{1-e_2^2}\cos{i_2} = \cos{i_*}$ leads to the inclination at the upper boundary \citep{naoz2016eccentric},
\begin{equation}\label{Eq15}
\cos{i_2} = \pm \sqrt{\frac{3}{5}}
\end{equation}
which gives $i_2 \approx 39.23^{\circ}$ or $i_2 \approx 140.77^{\circ}$ (called Kozai angles). This means that the boundary of inclination ($i_2$) for Kozai librations is independent on the Kozai parameter $i_*$. In Fig. \ref{Fig7_1}, the separatrices corresponding to $i = 39.23^{\circ}$ and $i = 140.77^{\circ}$ are shown in black lines. In addition, the conservation of $H$ means that the inclination of test particles cannot go over the limit of $i=90^{\circ}$, so that the Kozai oscillations are confined to the prograde region with inclination ranging from $39.23^{\circ}$ to $90^{\circ}$ or to retrograde region with inclination ranging from $90^{\circ}$ to $140.77^{\circ}$.

\begin{figure*}
\centering
\includegraphics[width=0.8\textwidth]{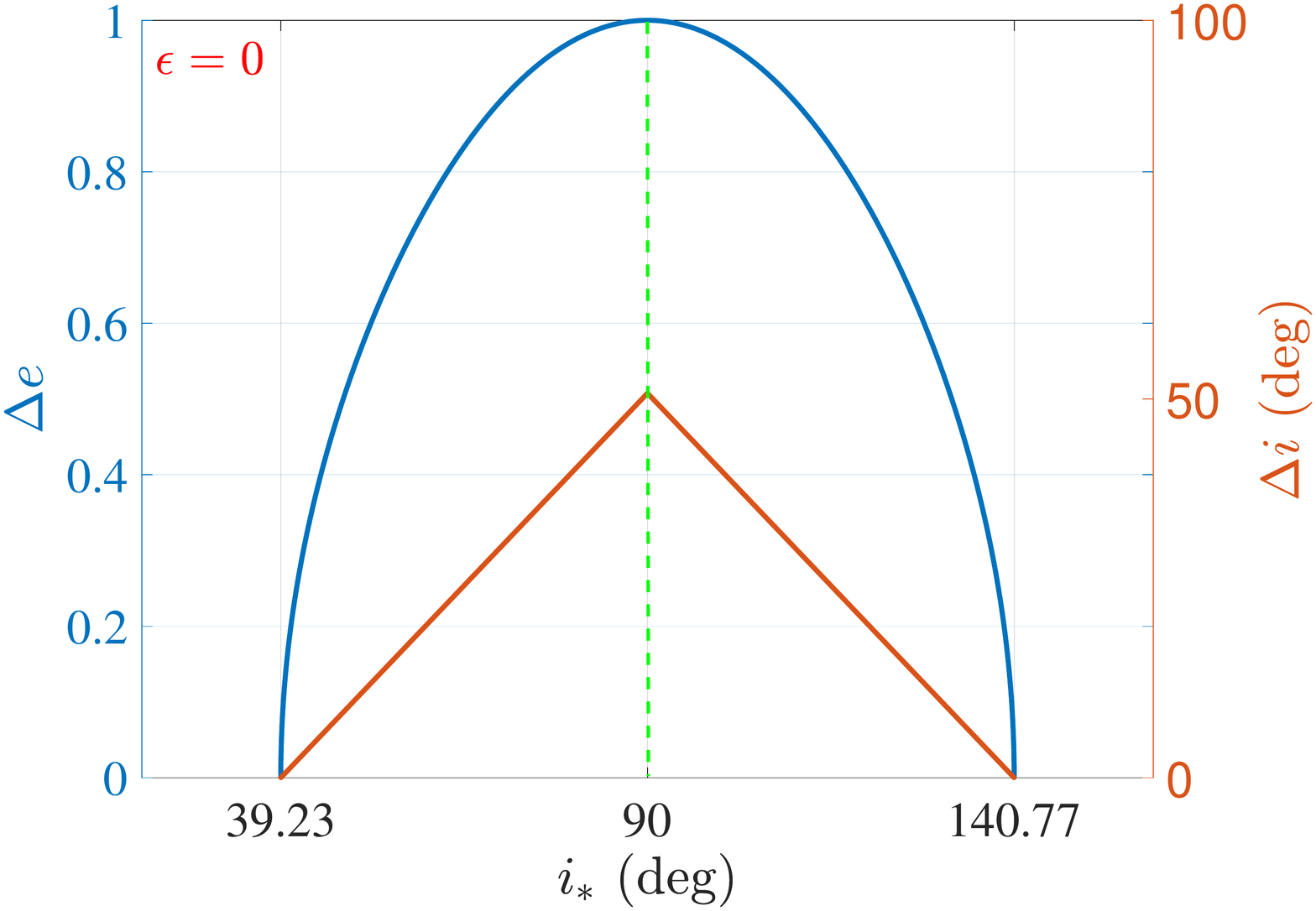}
\caption{Libration width in terms of $\Delta e$ and $\Delta i$ for Kozai resonance under the dynamical model at the quadrupole-level approximation. The vertical green dashed line stands for the location of $i_* = 90^{\circ}$.}
\label{Fig7_2}
\end{figure*}

Evidently, the resonant width in terms of the variation of eccentricity is $\Delta e = e_2 - e_1 = \sqrt{1-\frac{5}{3}\cos^2{i_*}}$ and the one in terms of the variation of inclination is $\Delta i = i_2 - i_1 = i_* - 39.23^{\circ}$ for the prograde case and $\Delta i = i_2 - i_1 = 140.77^{\circ} - i_*$ for the retrograde case. Please see Fig. \ref{Fig7_2} for $\Delta e$ and $\Delta i$ as functions of the Kozai parameter $i_*$. It is observed that, when the parameter $i_*$ is closer to $90^{\circ}$, the resonant width ($\Delta e$ or $\Delta i$) is larger.

In summary, at the quadrupole-level approximation ($\epsilon = 0$), the dynamical model is totally integrable. In the phase space with Kozai parameter $39.23^{\circ} < i_* < 140.77^{\circ}$, there is only Kozai resonance and the resonant centre is located at ($2g = \pi$, $G^2=\sqrt{5/3} \left|\cos{i_*}\right|$). For such an integrable model, the modes of motion are determined by the Hamiltonian ${\cal H}$ and motion integral $H$ (or $i_*$) and the separatrix between libration and circulation regions is given by the critical line of ${\cal H} = -H^2 = -\cos^2{i_*}$ (see Fig. \ref{Fig7}). Evaluating the resonant width of Kozai island, we can further get that the libration zone of Kozai resonance covers the inclination space ranging from $i = 39.23^{\circ}$ to $90^{\circ}$ (oscillations in the prograde region) and from $i = 90^{\circ}$ to $140.77^{\circ}$ (oscillations in the retrograde region), and the resonant width ($\Delta e$ or $\Delta i$) is larger when $i_*$ is closer to $90^{\circ}$, as shown in Figs. \ref{Fig7_1} and \ref{Fig7_2}.

\section{Poincar\'e surfaces of section}
\label{Sect3}

When the parameter $\epsilon$ is different from zero, the dynamical model is a two-degree-of-freedom system but with one conserved quantity (i.e., the Hamiltonian), leading to the fact that, in general, the resulting dynamical model is not integrable. For such a kind of two degree of freedom model, Poincar\'e surface of section is a powerful tool to uncover the global dynamics in the phase space \citep{lithwick2011eccentric, li2014chaos}. Thus, in this section, we will analyse the surfaces of section at different levels of Hamiltonian to get some preliminary dynamics.

In practice, we define two types of sections. In the first type, the section is defined as
\begin{equation*}
\Omega = \pi/2,\quad {\dot \Omega}  {{\dot \Omega}_0} > 0,
\end{equation*}
and, in the second type, the section is defined by
\begin{equation*}
\Omega = \pi,\quad {\dot \Omega}  {{\dot \Omega}_0} > 0.
\end{equation*}
The condition ${\dot \Omega}  {{\dot \Omega}_0} > 0$ implies that the test particle crosses the section of $\Omega = \pi/2$ (or $\Omega = \pi$) in the same direction (${\dot \Omega}_0$ stands for the initial time derivative of the longitude of ascending node).

It should be mentioned that our choice of section at $\Omega = \pi/2$ (the first type) and $\Omega = \pi$ (the second type) is based on the consideration that there is a correspondence between stable periodic orbits and centres of libration island arising in the Poincar\'e sections and thus the initial condition of stable periodic orbits can be provided by locating the centre of libration islands. In addition, it is better to place the starting points of symmetric periodic orbits at the aligned apsides with $\varpi_0 = 0$ or at the anti-aligned apsides with $\varpi_0 = 0$. As a result, the first-type section defined by $\Omega = \pi/2$ aims to produce periodic orbits associated with the resonances centred at $\omega = \pi/2$ or $\omega = 3\pi/2$ (for example, Kozai resonance) and the second-type section defined by $\Omega = \pi$ aims to generate those periodic orbits associated with the resonances centred at $\omega = \pi$ or $\omega = 0$ (for example, the apsidal resonance).

\begin{figure*}
\centering
\includegraphics[width=0.48\textwidth]{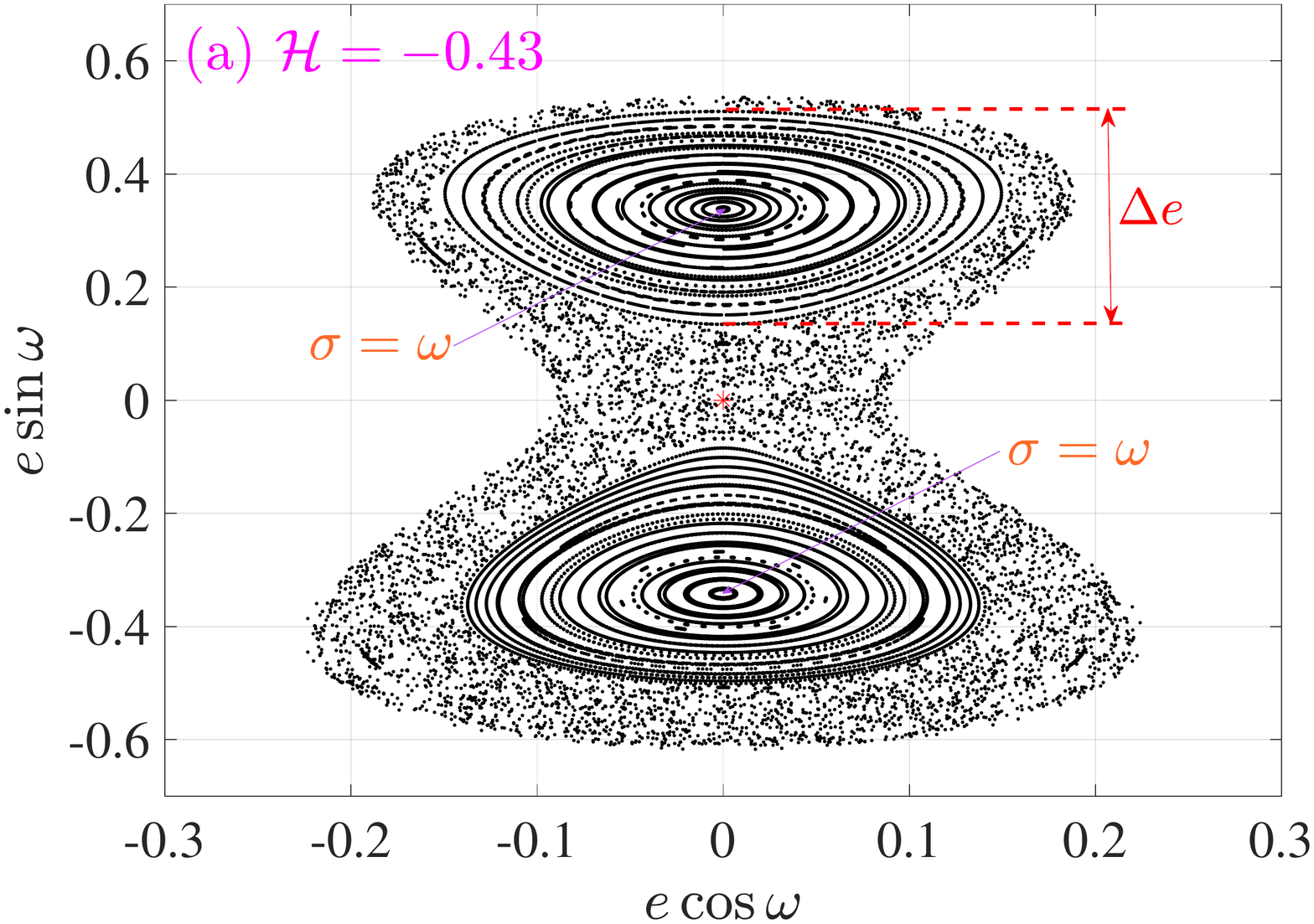}
\includegraphics[width=0.48\textwidth]{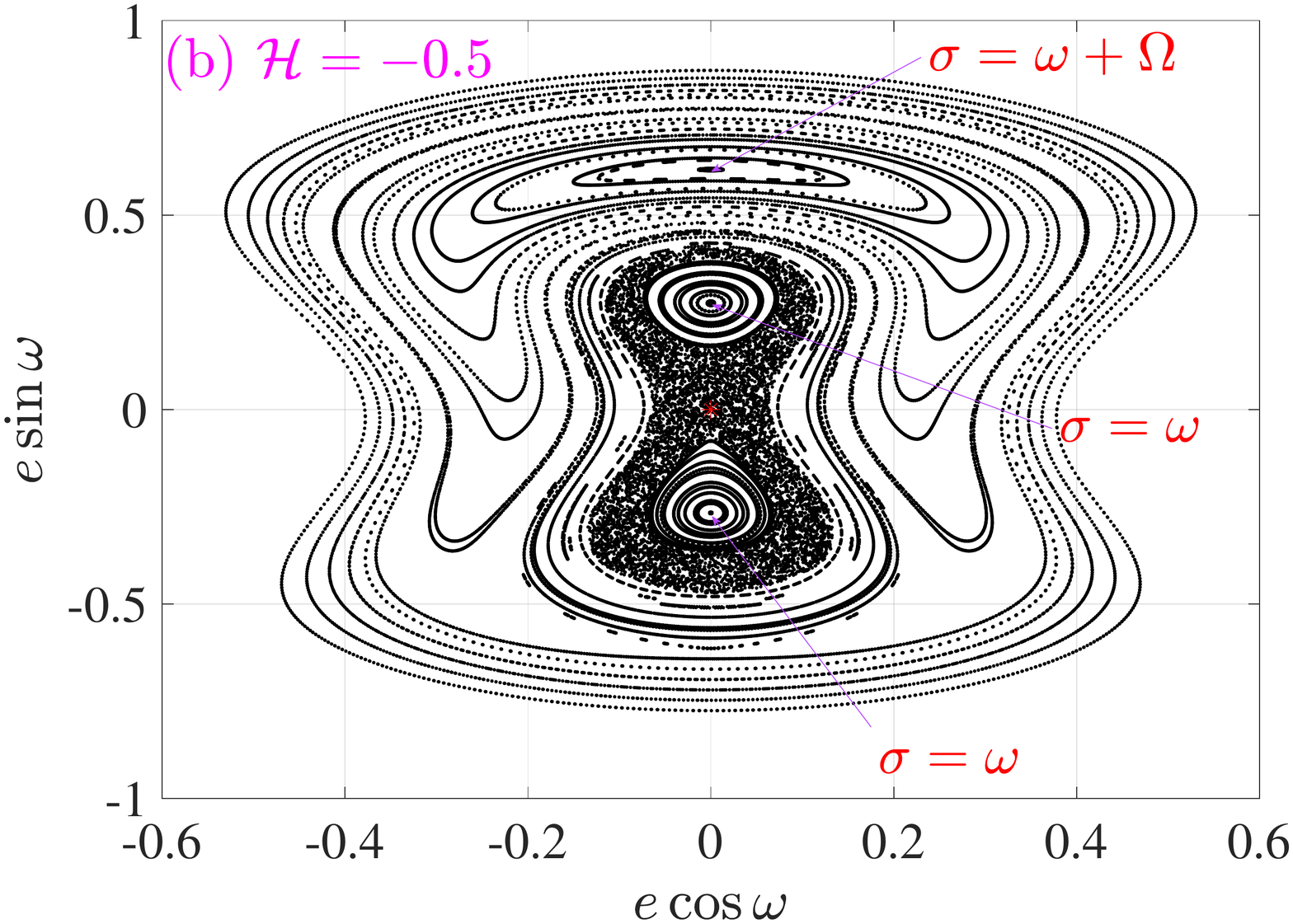}\\
\includegraphics[width=0.48\textwidth]{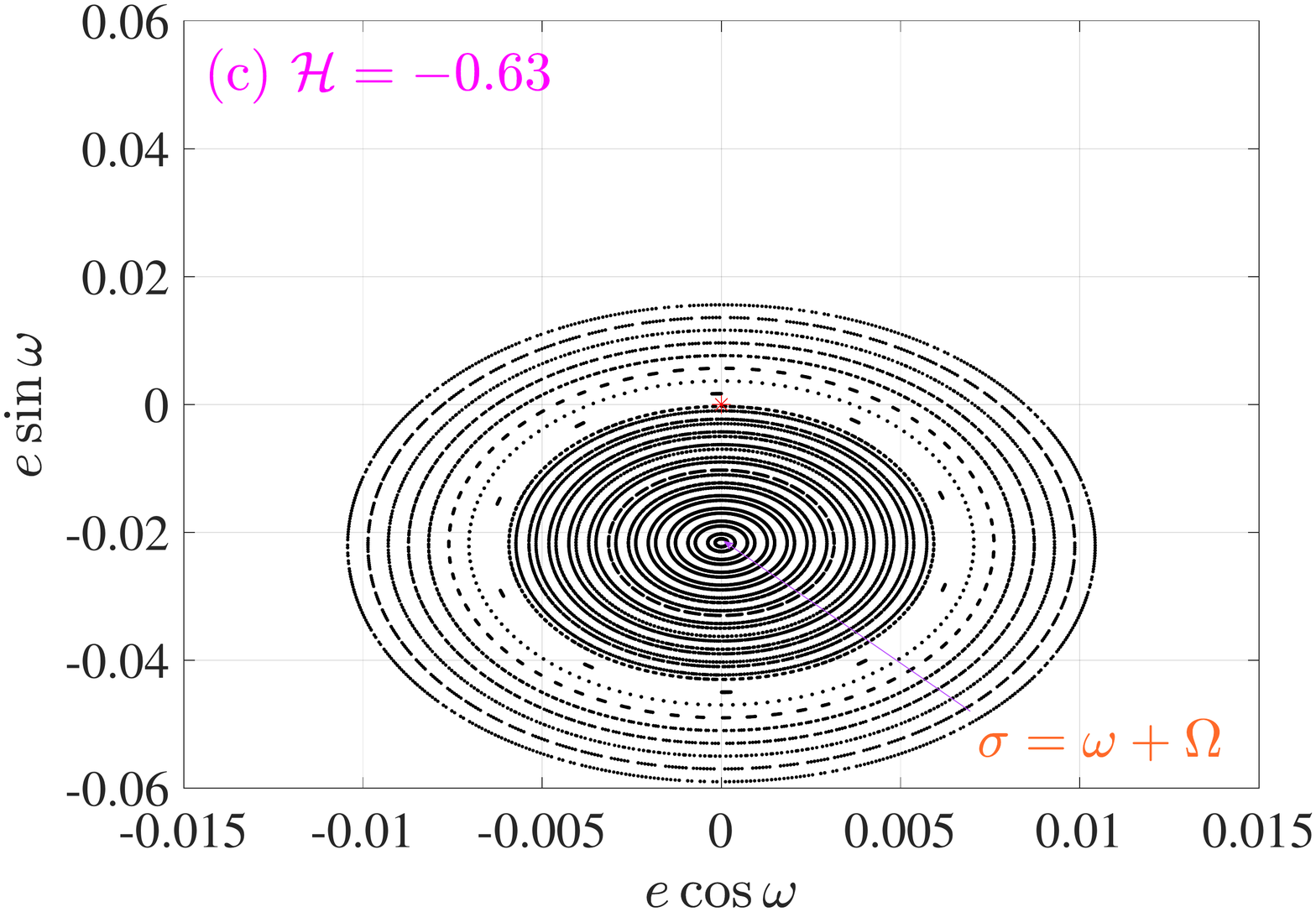}
\includegraphics[width=0.48\textwidth]{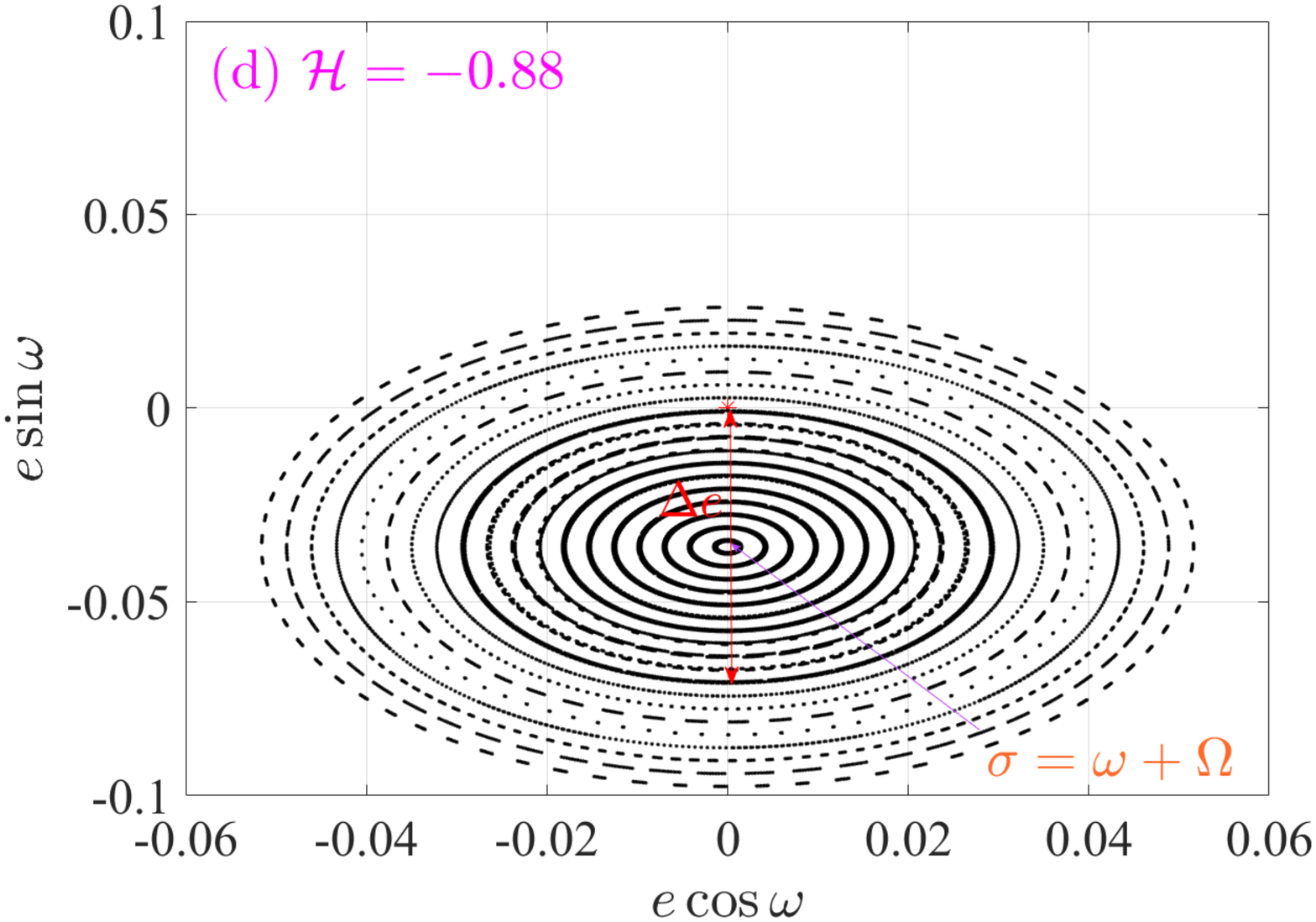}
\caption{Poincar\'e surfaces of section (defined by $\Omega = \pi/2$) for different levels of Hamiltonian. The red star represents the zero-eccentricity point. Note that all points in the Poincar\'e section are prograde.}
\label{Fig1}
\end{figure*}

The points on the defined sections are presented in the $(e\cos\omega, e\sin\omega)$ plane, in which it is clearly observed that the regular regions are filled with smooth curves and chaotic regions are filled by scattered points. In the regular regions, there are resonant and circulatory trajectories and, in the regions filled with scattered points, the motions are chaotic \citep{li2014chaos}. On the Poincar\'e sections, at each point specified by $(e\cos\omega,e\sin\omega)$, the variable $\dot \Omega$ can be retrieved from the given Hamiltonian ${\cal H}$. Thus, the Poincar\'e sections could provide the full state information for those trajectories crossing the defined sections for a given Hamiltonian value. For the simulations performed in this section, we take the factor $\epsilon = 0.03$ as an example to characterise the dynamical model. The Poincar\'e sections of the first type (defined by $\Omega = \pi/2$) are reported in Fig. \ref{Fig1} for different levels of Hamiltonian.

When the Hamiltonian is ${\cal H} = -0.43$ (see panel `a'), there are two islands of libration centred at $\omega = 90^{\circ}$ and $\omega = 270^{\circ}$. It is observed that these two islands are surrounded by chaotic regions, and the chaotic regions are distributed along the separatrix of the Kozai resonance shown in Fig. \ref{Fig6}. This phenomenon can be understood as follows: the perturbation transforms the separatrix into a chaotic layer, and such a chaotic layer becomes larger and larger when the perturbation increases.

In the centre of each island, there is a (stable) periodic orbit corresponding to the resonant centre with $\omega$ at $90^{\circ}$ and $270^{\circ}$ (these two islands of libration are due to the well-known Kozai resonance), and inside the island the motions are quasi-periodic trajectories (also known as resonant trajectories). The size of the libration island, marked by $\Delta e$, can measure the range of the stable libration zone of Kozai resonance, which is to be discussed in Sect. \ref{Sect5} for more details.

When the Hamiltonian is decreased to ${\cal H} = -0.5$, the structures of Poincar\'e sections are significantly changed (see panel `b'). Besides the islands appearing in panel (a), an additional island of libration arises at $\omega = 90^{\circ}$ in the region with higher eccentricities. Totally, there are three islands of libration in the considered section, two of them correspond to the Kozai resonance with argument of $\sigma = \omega$ (the resonant centres are located at $\sigma = 90^{\circ}$ and $\sigma = 270^{\circ}$), and the third island with a higher eccentricity is due to the (anti-aligned) apsidal secular resonance with critical argument $\sigma = \omega + \Omega$ librating around $\sigma = \pi$.

When the Hamiltonian is further decreased to ${\cal H} = -0.63$ (see panel `c') and ${\cal H} = -0.88$ (see panel `d'), the dynamical structures arising in the Poincar\'e section become relatively simple. In both panels, there is only one island of libration in the section, which corresponds to the (aligned) apsidal secular resonance with $\sigma = \omega + \Omega$ librating around $\sigma = 0$. The size of the libration island is also marked by $\Delta e$, which could measure the range of stable libration zone, as shown in panel (d).

From Fig. \ref{Fig1}, we can see that there exists Kozai resonance in the Poincar\'e section when the Hamiltonian is equal to ${\cal H} = -0.43, -0.5$ and there is no Kozai resonance when the Hamiltonian is equal to ${\cal H} = -0.63, -0.88$. This is in agreement with the discussion made in Section \ref{Sect2_2} that the occurrence of Kozai resonance requires the Hamiltonian $-0.6 < {\cal H} < 3$. However, the perturbation due to the octupole-order terms in the Hamiltonian introduces significant changes about the dynamical structures. To show this point, here we make a direct comparison between the phase portraits produced under the quadrupole-level approximation (see Fig. \ref{Fig6}) and the Poincar\'e surfaces of section obtained from the octupole-order approximation (see Fig. \ref{Fig1}). Firstly, the comparison between panel (a) of Fig. \ref{Fig6} and panels (c,d) of Fig. \ref{Fig1} shows that there are no any resonance under the quadrupole-order model but, under the octupole-order model, the apsidal secular resonance with critical argument $\sigma = \omega + \Omega$ appears. Secondly, the comparison between panel (b) of Fig. \ref{Fig6} and panels (a,b) of Fig. \ref{Fig1} indicates that (a) the separatrix appearing in the phase portrait under the quadrupole-order model is replaced by a chaotic layer under the octupole-order model and (b) besides the Kozai resonance the apsidal secular resonance with critical argument of $\sigma = \omega + \Omega$ takes place under the octupole-order model.

\begin{figure*}
\centering
\includegraphics[width=0.48\textwidth]{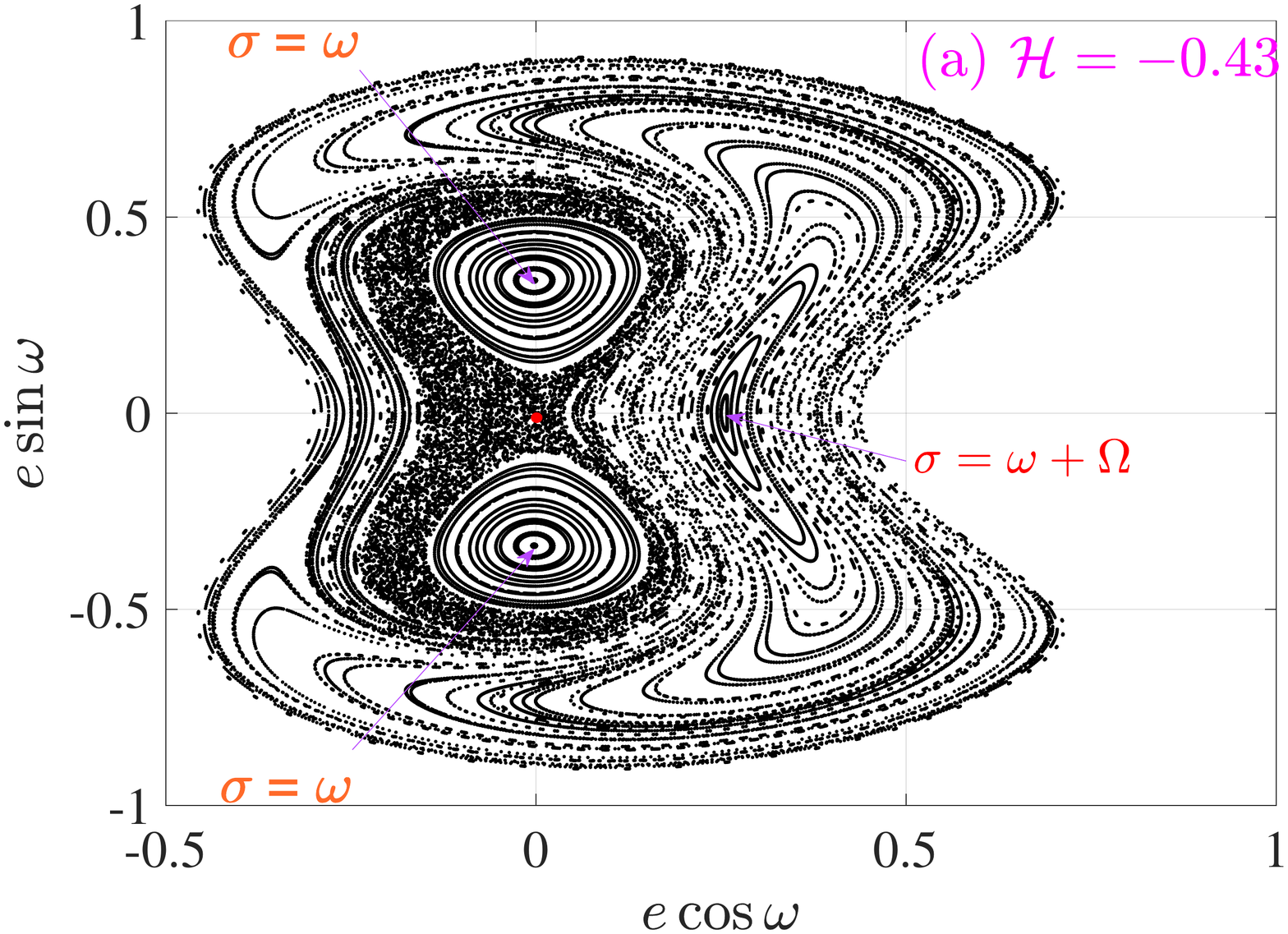}
\includegraphics[width=0.48\textwidth]{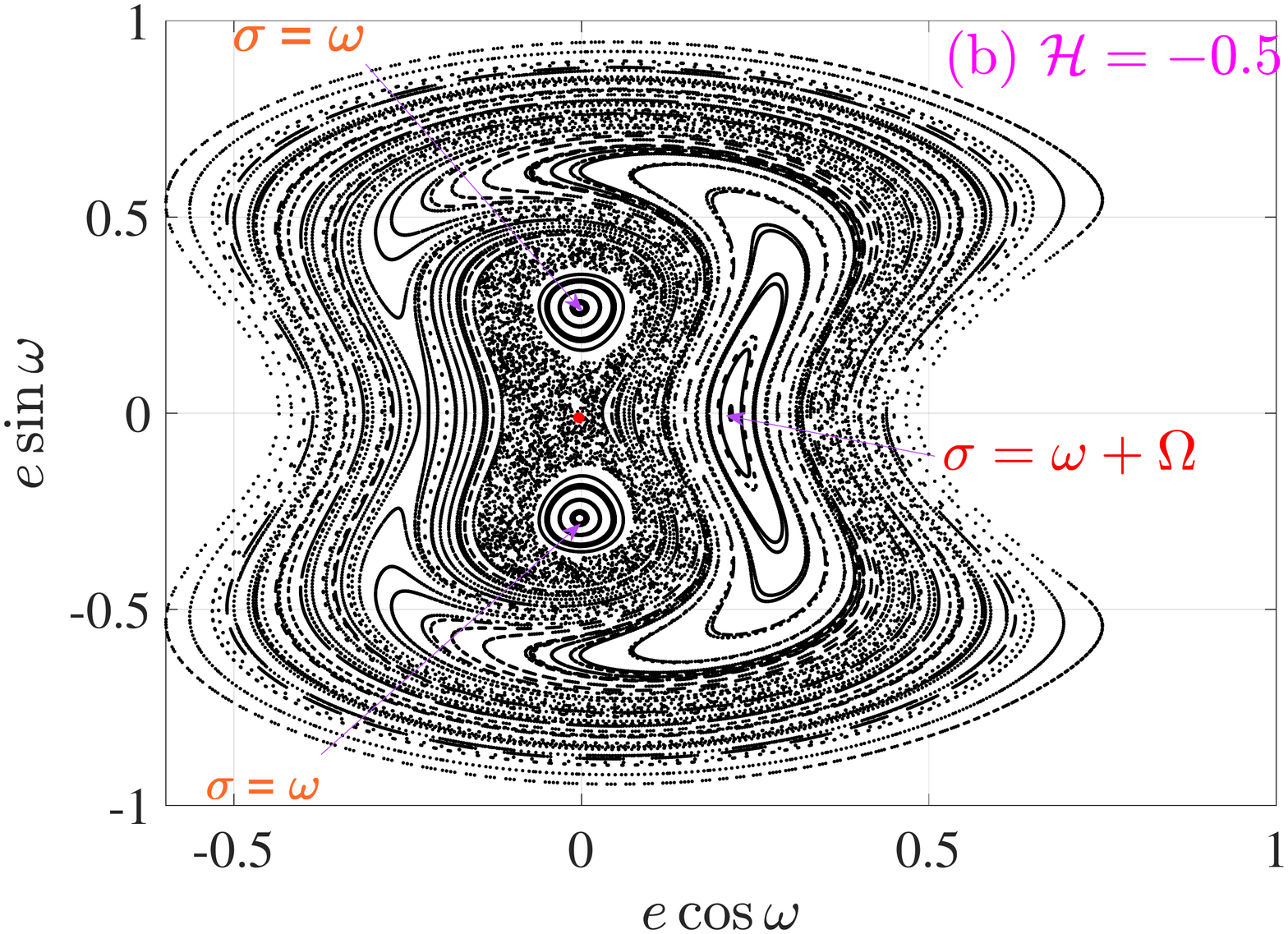}
\caption{Poincar\'e surfaces of section (defined by $\Omega = \pi$) for two levels of Hamiltonian at ${\cal H}=-0.43$ and ${\cal H}=-0.5$. The red dot stands for the zero-eccentricity point. Note that all points in the Poincar\'e section are on prograde orbits.}
\label{Fig2}
\end{figure*}

In Fig. \ref{Fig2}, the Poincar\'e sections of the second type (defined by $\Omega = \pi$) are reported. The Hamiltonian is taken as ${\cal H} = -0.43$ in the left-hand panel and ${\cal H} = -0.5$ in the right-hand panel. In each panel, it is observed that there are three islands of libration. The first two islands centred at $\omega = 90^{\circ}$ and $\omega = 270^{\circ}$ corresponds to the Kozai resonance, which have been appeared in the first two panels of Fig. \ref{Fig1}. The third island of libration centred at $\omega = 0^{\circ}$ is a new one in comparison to the ones appearing in Fig. \ref{Fig1} and it is due to the (anti-aligned) apsidal secular resonance with $\sigma = \omega + \Omega$ librating around $\sigma = \pi$.

At the centre of each island of libration, there is a (stable) periodic orbit. According to dynamical system theory, the (stable) periodic orbit projected onto the Poincar\'e section corresponds to the resonant centre, and the quasi-periodic orbits correspond to the associated librating trajectories inside the islands \citep{lithwick2011eccentric, li2014chaos, naoz2016eccentric}. Thus, we could compute stable periodic orbits in order to identify the location of resonant centres and then evaluate the size of libration islands filled with quasi-periodic orbits in order to specify the boundaries of stable libration zones. This is the key idea of the current work.

In the current study, we concentrate on the symmetric periodic orbits with starting points on the aligned or anti-aligned apsides (i.e., $\varpi_0 = 0$ or $\varpi_0 = \pi$). Those asymmetric periodic orbits and their associated resonances are out of the scope of the current study. According to the Poincar\'e sections shown in Fig. \ref{Fig1}, we can summarise that the initial angles of the first two types of symmetric periodic orbits are $\omega_0 = 3\pi/2, \Omega_0 = \pi/2$ (case I) and $\omega_0 = \pi/2, \Omega_0 = \pi/2$ (case II), and the sections shown in Fig. \ref{Fig2} indicate that the initial angles of the third type of symmetric periodic orbits are $\omega_0 = 0$ and $\Omega_0 = \pi$ (case III).

It should be mentioned that this study is restricted to those major secular resonances that are visible in the Poincar\'e surfaces of section, and thus they only represent a subset of varieties of secular resonances. This choice is justified by the fact that these resonances considered are the strongest ones, dominating the secular dynamics of test particles.

In the coming section, we determine the initial eccentricity and inclination for symmetric periodic orbits of cases I, II and III (their initial angles $\Omega_0$ and $\omega_0$ are known).

\section{Webs of secular resonances (or families of stable periodic orbits)}
\label{Sect4}

As discussed in the previous section, the appearance of libration islands in the Poincar\'e sections is due to secular resonances and there is a correspondence between stable periodic orbits and resonant centres. Thus, we could determine the exact location of resonance by computing the associated (stable) periodic orbit.

In the two-degree-of-freedom system, a periodic orbit should satisfy the following constraints:
\begin{equation*}
{\left\{ {k,h,q,p} \right\}_{t = 0}} = {\left\{ {k,h,q,p} \right\}_{t = T}},
\end{equation*}
where $T$ is the period. Obviously, the topic of solving periodic orbits is a typical two-point boundary-value problem (TPBVP). In general, for such a TPBVP, we need to determine five parameters including the initial states $(k_0,h_0,q_0,p_0)$ and period $T$.

Let us take case I as an example. The initial states of periodic orbits (for case I) can be expressed as follows (it is known that the initial angles are $\omega_0 = 3\pi/2$ and $\Omega_0 = \pi/2$):
\begin{equation*}
k_0 = 0,\quad h_0 = e_0,\quad q_0 = 0,\quad p_0 = \sin{\frac{i_0}{2}},
\end{equation*}
where only the initial eccentricity and inclination ($e_0$ and $i_0$) are undetermined variables. In addition, for a given Hamiltonian ${\cal H}$, only one of the variables $e_0$ and $i_0$ is independent, i.e., $e_0$ (or $i_0$) can be known from the expression of Hamiltonian ${\cal H} (e_0, i_0, \Omega_0 = 3\pi/2, \omega_0 = \pi/2)$.
In summary, for a given Hamiltonian, we only need to identify two variables including $e_0$ (or $i_0$) and $T$ by means of correction algorithm. The Hamiltonian ${\cal H}$ varies in a certain interval, then the entire family of periodic orbits can be obtained by means of continuation algorithm. Please refer to \citet{howell1984three} and \citet{lara2002numerical} for the correction and continuation algorithms.

To judge the stability of a periodic orbit, we need to know its Monodromy matrix, which corresponds to the state transition matrix (STM) evaluated at one period. For the two-degree of freedom dynamical model considered in the present study, there are two pairs of eigenvalues for the Monodromy matrix. One pair of them is equal to unity, and the other non-unity pair of eigenvalues determine the stability. Let us denote the non-unity pair of eigenvalues as $(\lambda_1,\lambda_2)$ and define the stability index as $S = \lambda_1 + \lambda_2$. The stable condition of periodic orbits states that the stability index $S$ should be smaller than 2, i.e., $S < 2$ \citep{henon1969numerical}.

\begin{figure*}
\centering
\includegraphics[width=0.8\textwidth]{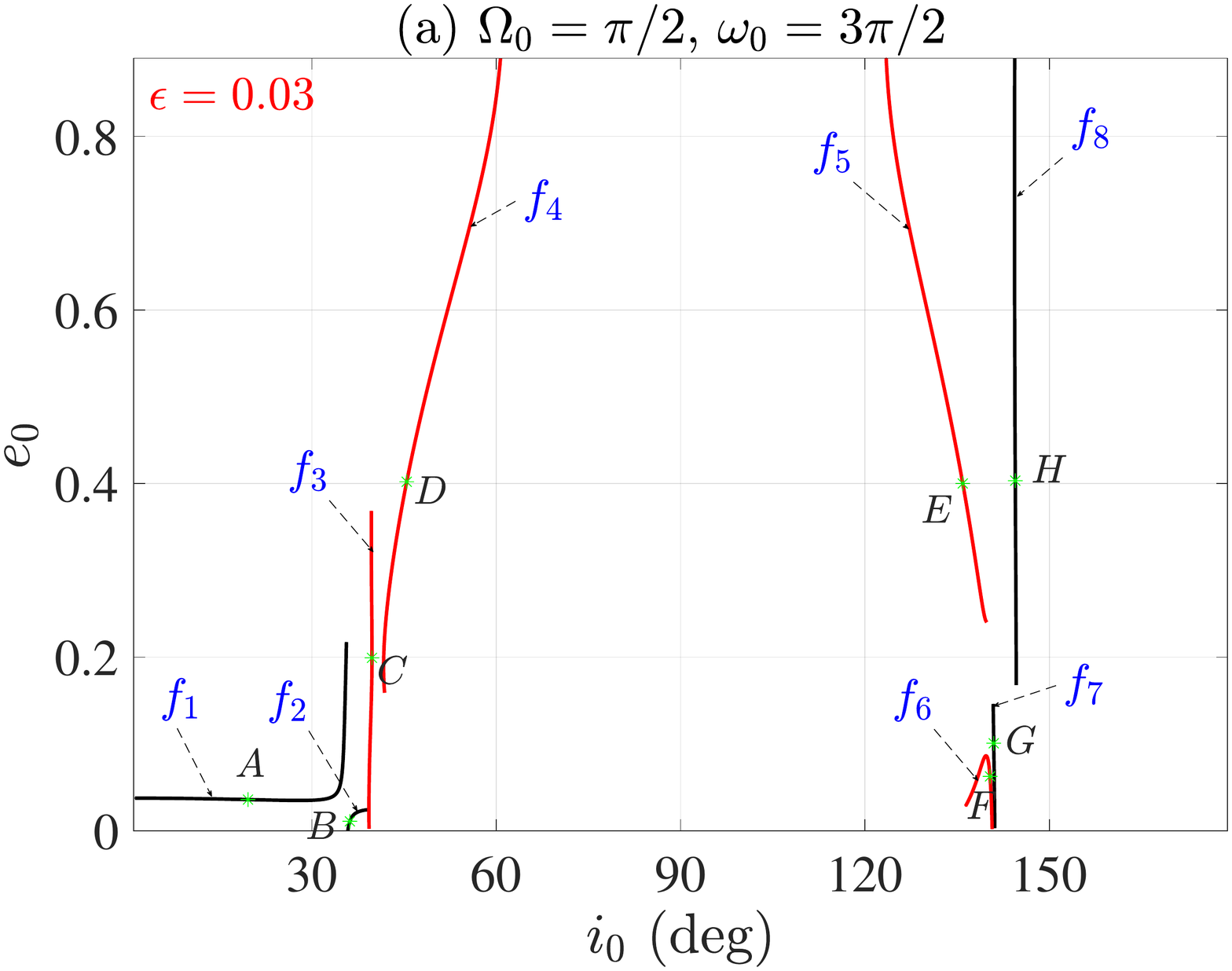}
\includegraphics[width=0.8\textwidth]{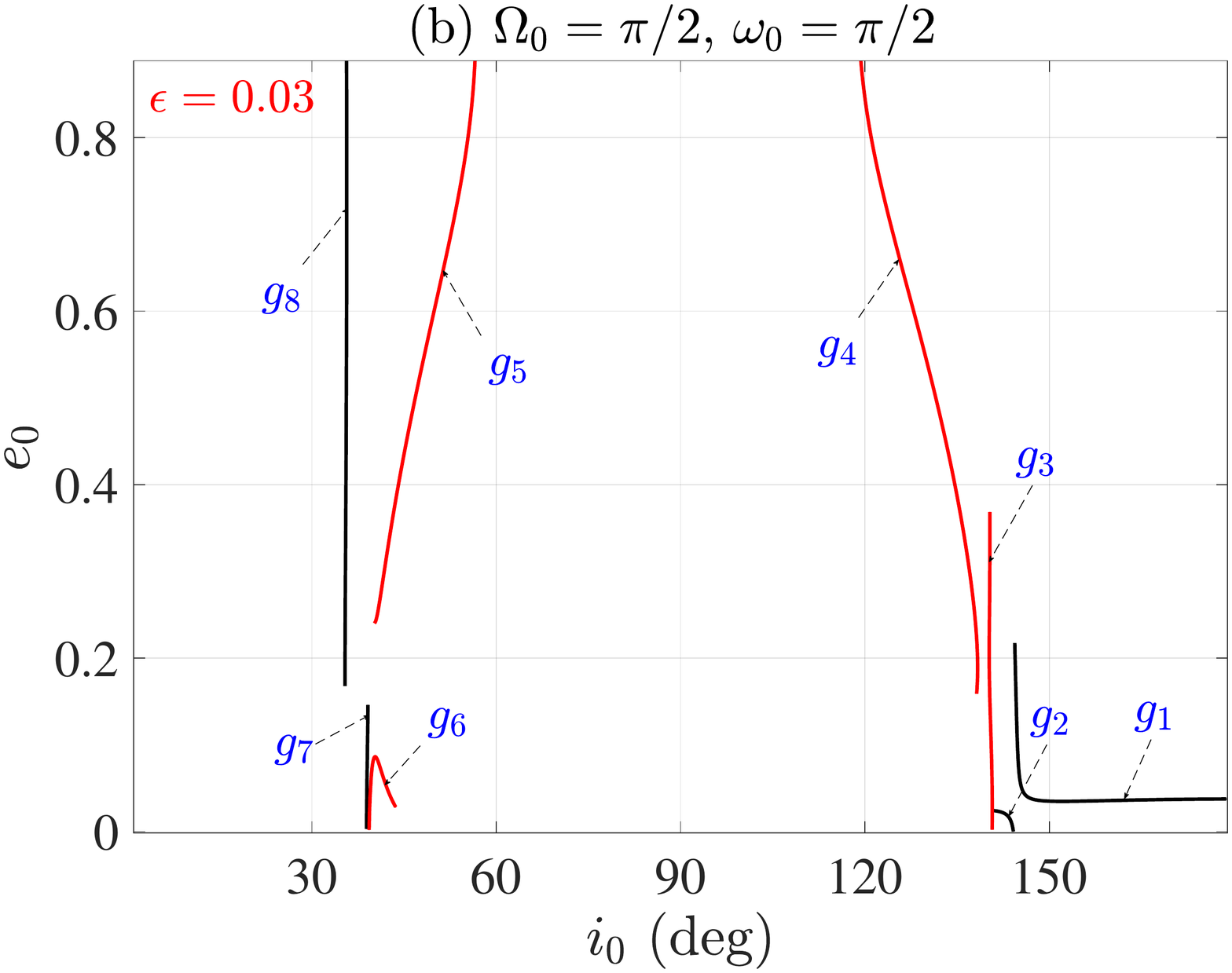}
\includegraphics[width=0.8\textwidth]{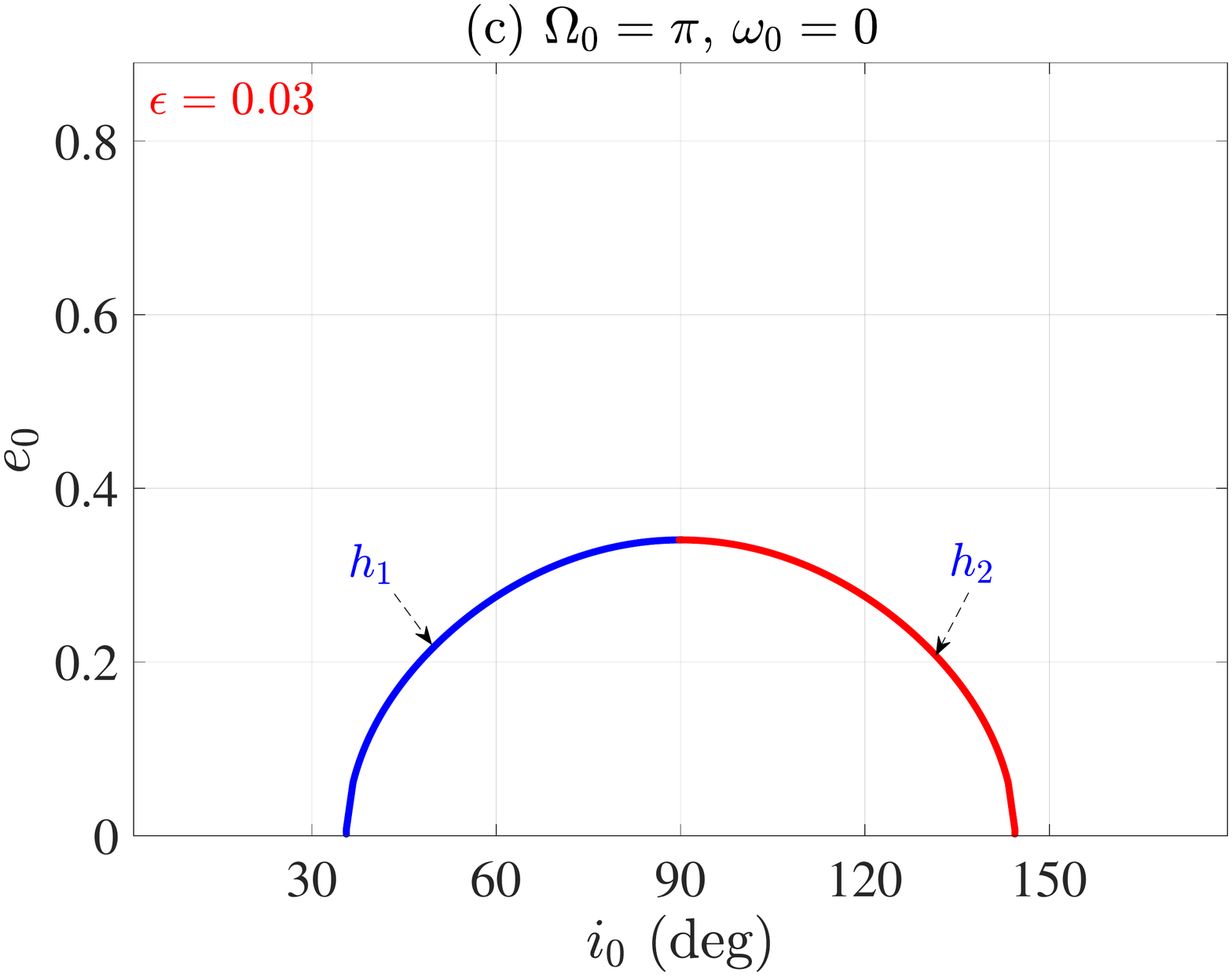}
\caption{Characteristic curves of families of stable periodic orbits (or webs of secular resonances) in the case of $\epsilon = 0.03$ for cases I, II and III.}
\label{Fig3}
\end{figure*}

In Fig. \ref{Fig3}, the characteristic curves of stable periodic orbits ($S < 2$) are presented in the action space $(e_0, i_0)$ for cases I, II and III. In case I (see panel a) and case II (see panel b), there are eight families of periodic orbits in the whole action space, and they are denoted by $f_i$ for case I and $g_i$ for case II ($i=1,2,...,8$). In case III (see panel c), there are two families of stable periodic orbits, which are denoted by $h_1$ for the prograde case and $h_2$ for the retrograde case. In a certain family of periodic orbits, all the members hold the same resonant argument. Please refer to Table \ref{Tab0} for the resonant arguments and their libration centres.

\begin{table}
\centering
\caption{Resonant arguments for the families of stable periodic orbits shown in Fig. \ref{Fig3}. $\sigma_i$ stands for the critical argument and $\sigma_{i,c}$ is the associated resonant centre.}
\begin{tabular*}{\hsize}{@{}@{\extracolsep{\fill}}lccc@{}}
\hline
Family & Resonances & $\sigma_i$ & $\sigma_{i,c}$\\
\hline\hline
$f_1 (g_1)$& Aligned apsidal secular resonance & $\omega + \Omega$ ($\omega - \Omega$) & $0^{\circ}$\\
$f_2 (g_2)$& Aligned apsidal secular resonance & $\omega + \Omega$ ($\omega - \Omega$)  & $0^{\circ}$\\
$f_3 (g_3)$& Kozai resonance& $\omega$ & $270^{\circ}$ ($90^{\circ}$)\\
$f_4 (g_4)$& Kozai resonance& $\omega$ & $270^{\circ}$ ($90^{\circ}$)\\
$f_5 (g_5)$& Kozai resonance& $\omega$ & $270^{\circ}$ ($90^{\circ}$)\\
$f_6 (g_5)$& Kozai resonance& $\omega$ & $270^{\circ}$ ($90^{\circ}$)\\
$f_7 (g_7)$& High-order secular resonance& $3\omega - \Omega$ ($3\omega + \Omega$) & $0^{\circ}$\\
$f_8 (g_8)$& Anti-aligned apsidal secular resonance& $\omega - \Omega$ ($\omega + \Omega$)  & $180^{\circ}$\\
$h_1 (h_2)$& Anti-aligned apsidal secular resonance& $\omega + \Omega$ ($\omega - \Omega$)  & $180^{\circ}$\\
\hline
\end{tabular*}
\label{Tab0}
\end{table}

Observing the characteristic curves shown in Fig. \ref{Fig3}, we can find that (a) the members in $f_i$ and the ones in $g_i$ (with the same $i$) are symmetric with respect to the polar line and (b) the members in family $h_1$ and the ones in $h_2$ are symmetric relative to the polar line. These symmetric properties of solution are due to the symmetry inherent in the Hamiltonian, as discussed in Sect. \ref{Sect2}.

\begin{figure*}
\centering
\includegraphics[width=0.98\textwidth]{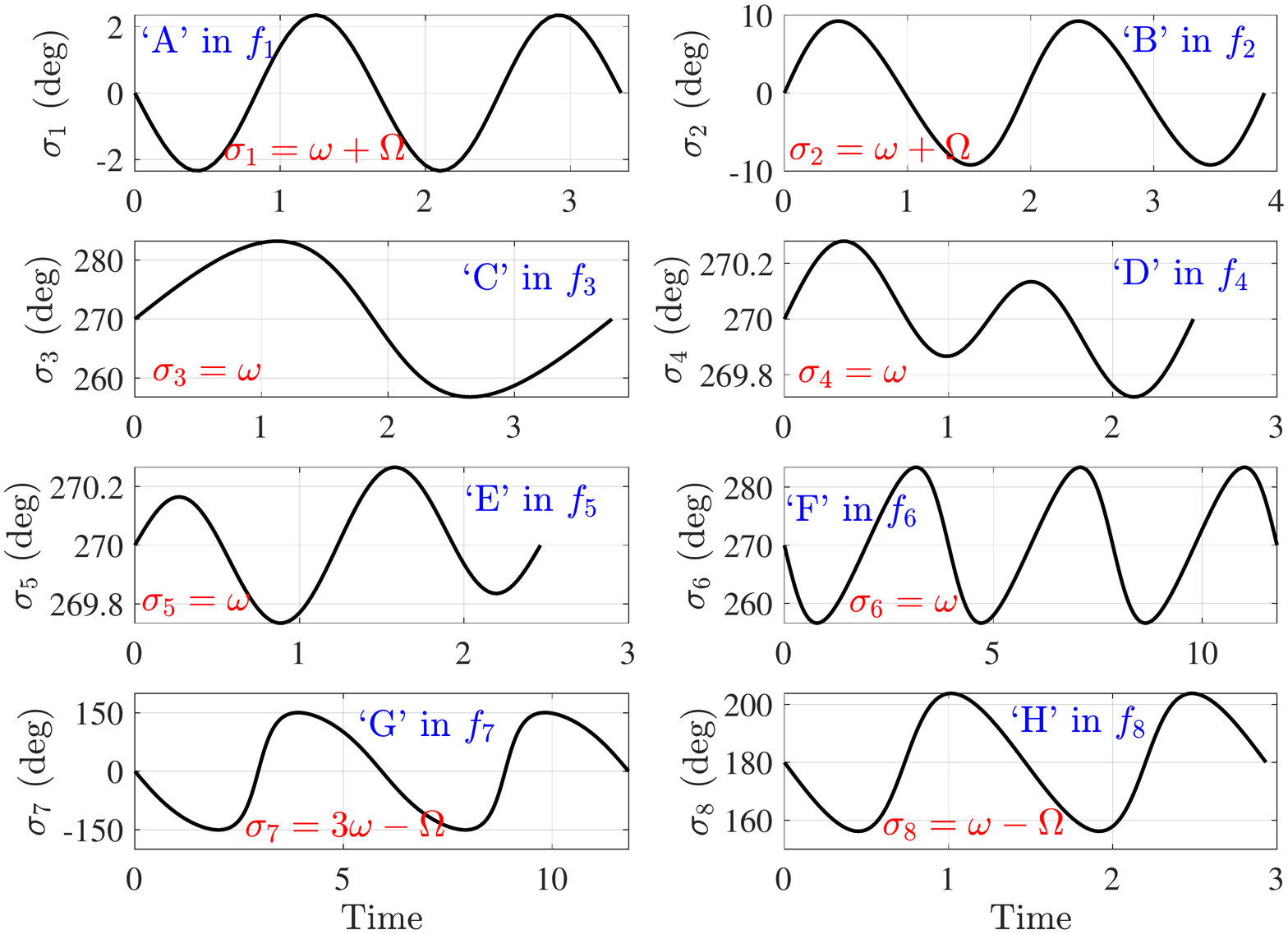}
\caption{Time histories of the resonant angle for the example periodic orbits marked in panel (a) of Fig. \ref{Fig3}.}
\label{Fig4}
\end{figure*}

In panel (a) of Fig. \ref{Fig3}, one representative member in each family is marked by a green star. For convenience, these representatives are denoted by letters from `A' in family $f_1$ to `H' in family $f_8$. Their time histories of resonant argument during one resonant period are shown in Fig. \ref{Fig4}. It is observed from Fig. \ref{Fig4} that (a) for members `A' in family $f_1$ and `B' in $f_2$, the resonant angle $\sigma_{1,2} = \omega + \Omega$ librates around $0^{\circ}$, (b) for members `C', `D', `E' and `F' in families $f_i$ with $i=3,4,5,6$, the resonant angle $\sigma_{i} = \omega$ librates around $270^\circ$, (c) for member `G' in family $f_7$, the resonant angle $\sigma_7 = 3\omega - \Omega$ librates around $0^{\circ}$ and (d) for member `H' in family $f_8$, the argument $\sigma_8 = \omega - \Omega$ librates around $180^{\circ}$.

From the viewpoint of geometrical configurations, we can see that the families of stable periodic orbits shown in Fig. \ref{Fig3} belong to several types of secular resonances (see Table \ref{Tab0}): (a) families $f_{1,2} (g_{1,2})$ correspond to the aligned apsidal secular resonances with critical argument librating around zero, (b) families $f_i (g_i)$ with $i=3,4,5,6$ correspond to Kozai resonances with $\omega$ as the critical argument librating around $90^{\circ}$ or $270^{\circ}$, (c) families $f_7(g_7)$ correspond to high-order secular resonances and (d) families $f_8 (g_8)$ and $h_1(h_2)$ correspond to the anti-aligned apsidal secular resonances with critical argument librating around $180^{\circ}$.

\begin{figure*}
\centering
\includegraphics[width=0.8\textwidth]{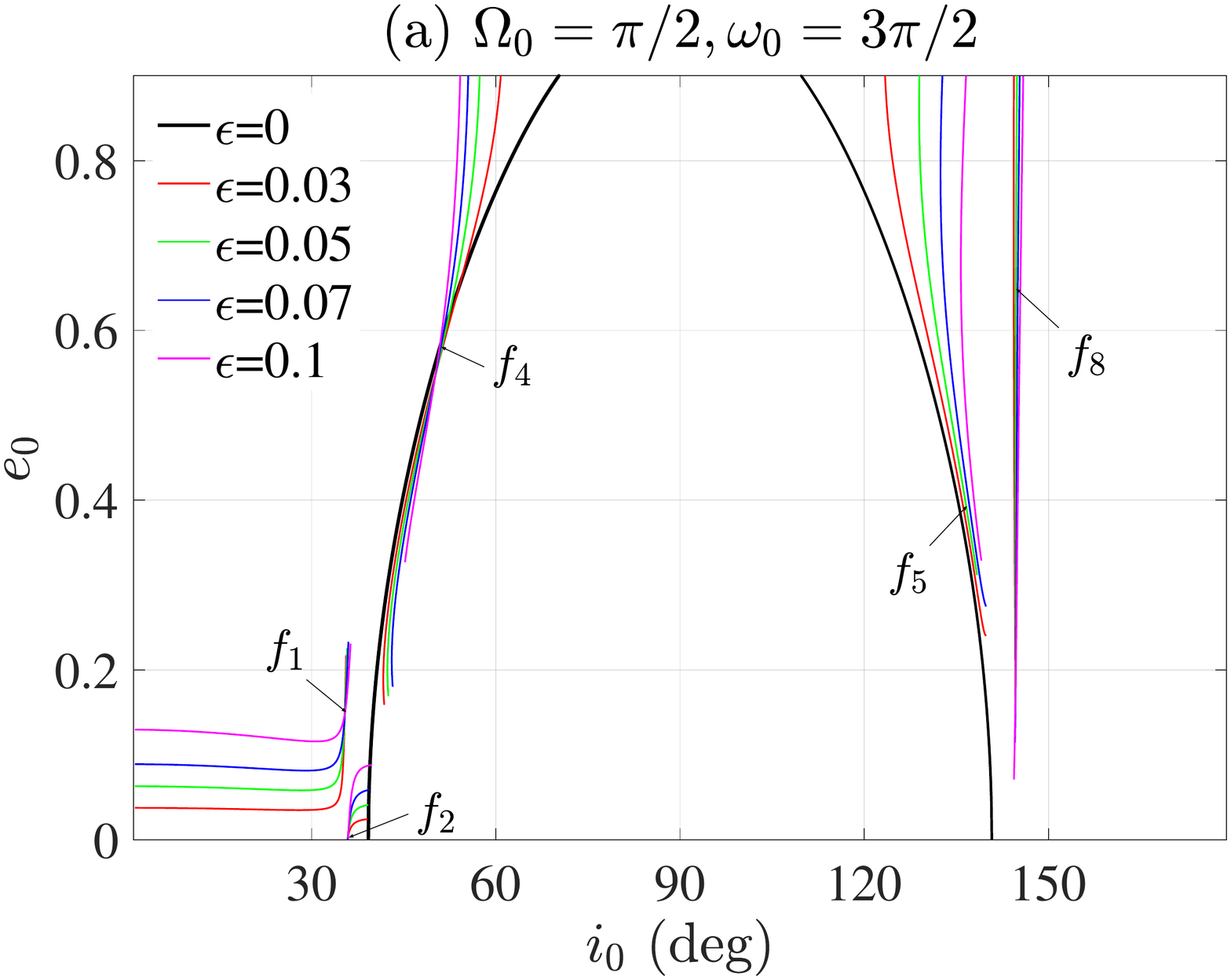}
\includegraphics[width=0.8\textwidth]{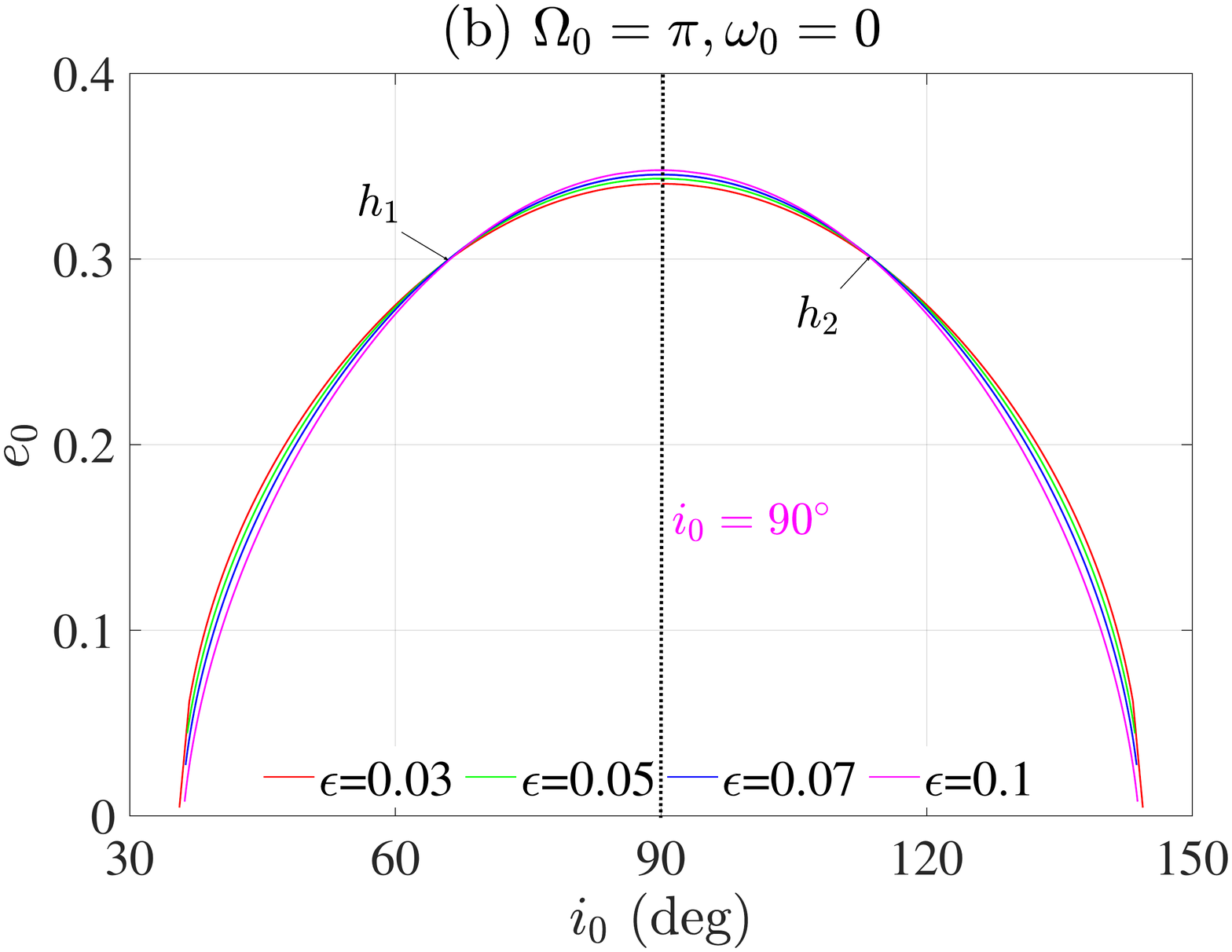}
\caption{Characteristic curves of stable periodic orbits (i.e., webs of secular resonances) obtained under dynamical models specified by different values of $\epsilon$. The upper panel is for case I and the bottom panel is for case III. The results in case II can be easily obtained by symmetry.}
\label{Fig9}
\end{figure*}

Following the similar approach, we can further compute families of stable periodic orbits under the dynamical models characterised by different values of $\epsilon$. The characteristic curves (or webs of secular resonances) of those primary families are reported in Fig. \ref{Fig9}. It is noted that families of stable periodic orbits in case II are not considered due to symmetry and families $f_3$, $f_6$ and $f_7$ are not presented considering their weak strength of resonance (their islands are not visible from Poincar\'e sections). For convenience of comparison, characteristic curves corresponding to $\epsilon = 0$ (see Fig. \ref{Fig7}) and the ones corresponding to $\epsilon = 0.03$ (see Fig. \ref{Fig3}) are also presented in Fig. \ref{Fig9}.

Observing Fig. \ref{Fig9}, we can see that (a) in the case of $\epsilon = 0$ (quadrupole-level approximation) there are only two families of stable periodic orbits corresponding to Kozai resonances (one is in the prograde region and the other one is in the retrograde region) and other families begin to appear when $\epsilon$ is different from zero, (b) characteristic curves are changed when $\epsilon$ is different, meaning that the location of resonant centres is dependent on the magnitude of $\epsilon$, (c) when $\epsilon$ is different from zero the characteristic curves of the families of Kozai resonance (i.e., families $f_4$ and $f_5$)  can not extend to the region around the zero-eccentricity point because the low-eccentricity regions are chaotic in dynamics, (d) in the case of $\epsilon=0$ (quadrupole-level approximation) the curves of Kozai centres are symmetric with respect to $i_0 = 90^{\circ}$, while such a symmetry breaks when $\epsilon$ is different from zero, (e) the characteristic curve of $f_8$ is insensitive to the eccentricity, (f) characteristic curves of $h_1$ and $h_2$ are always symmetric with respect to $i_0 = 90^{\circ}$ at arbitrary $\epsilon$ and (g) secular resonances in families $h_1$ and $h_2$ occur in the region with eccentricities smaller than $\sim$0.35.

In the next section, we will measure the boundaries of stable libration zones in the action space by analysing Poincar\'e sections.

\section{Boundaries of stable libration zones}
\label{Sect5}

In the previous section, we have produced the distribution of exact resonant centres in the action space $(e_0,i_0)$ by determining families of symmetric periodic orbits. In addition, we know that the islands of libration appearing in the Poincar\'e sections are due to the secular resonances, thus the boundaries of stable libration zones can be measured by analysing Poincar\'e sections.

In this work, we evaluate boundaries of stable libration zones when the angles are fixed at the resonant centre, i.e. fixing $\Omega_0 = \pi/2$ and $\omega_0 = 3\pi/2$ for case I (corresponding to families $f_i$ with $i=1,2,...,8$), $\Omega_0 = \pi/2$ and $\omega_0 = \pi/2$ for case II (corresponding to families $g_i$ with $i=1,2,...,8$), and $\Omega_0 = \pi$ and $\omega_0 = 0$ for case III (corresponding to families $h_1$ and $h_2$). It should be noted that the results (boundaries of stable libration zones) will be changed if we adopt different definitions of Poincar\'e sections.

To describe the process of evaluating libration zones in the space $(e_0,i_0)$, let us take case I ($\Omega_0 = \pi/2$ and $\omega_0 = 3\pi/2$) as an example. A procedure is designed to identify the boundaries of stable libration zones by searching for the upper and lower limits that enclose the islands of libration as follows:
\begin{enumerate}
 \item For a given Hamiltonian ${\cal H}_0$, the eccentricity and inclination ($e_0$ and $i_0$) of the exact resonant centre (i.e. the periodic orbit) can be determined, as shown in Figs \ref{Fig3} and \ref{Fig9}.
 \item Starting from $e_0$, increase the eccentricity and keep the Hamiltonian as a constant equal to ${\cal H}_0$. When the trajectory switches between libration and circulation, we halve the step of eccentricity until the step is smaller than a given tolerance. Let us record the maximum eccentricity at which the trajectory is of libration as $e_{\max}$.
 \item Similar to the previous step, decreasing the eccentricity from $e_0$ and keeping the Hamiltonian as a constant equal to ${\cal H}_0$, we can determine the minimum eccentricity at which the trajectory is of libration. Record the minimum value of eccentricity at which the trajectory is of libration as $e_{\min}$.
 \item According to the Hamiltonian ${\cal H}_0$, it is possible to reproduce the boundary points in the action space as $(e_{\min}, i_1)$ and $(e_{\max}, i_2)$.
 \item When the Hamiltonian varies, the points $(e_{\min},i_1)$ and $(e_{\max}, i_2)$ constitute the boundaries of stable libration zones.
\end{enumerate}

In practical simulations, we limit the duration of numerical integration as 1000 in normalised units, which is much longer than the timescales of secular resonances considered in the present work. In the regular regions, there is no problem to distinguish the points with libration and circulation. However, in the chaotic region, it becomes complicated and the following rules are adopted to distinguish orbits of libration and circulation: (a) those trajectories are recorded as librating orbits if their critical arguments are always of libration during the maximum period of time, and (b) those trajectories are recorded as non-libration if their critical arguments enter into the circulation mode in the process of numerical integration.

\begin{figure*}
\centering
\includegraphics[width=0.8\textwidth]{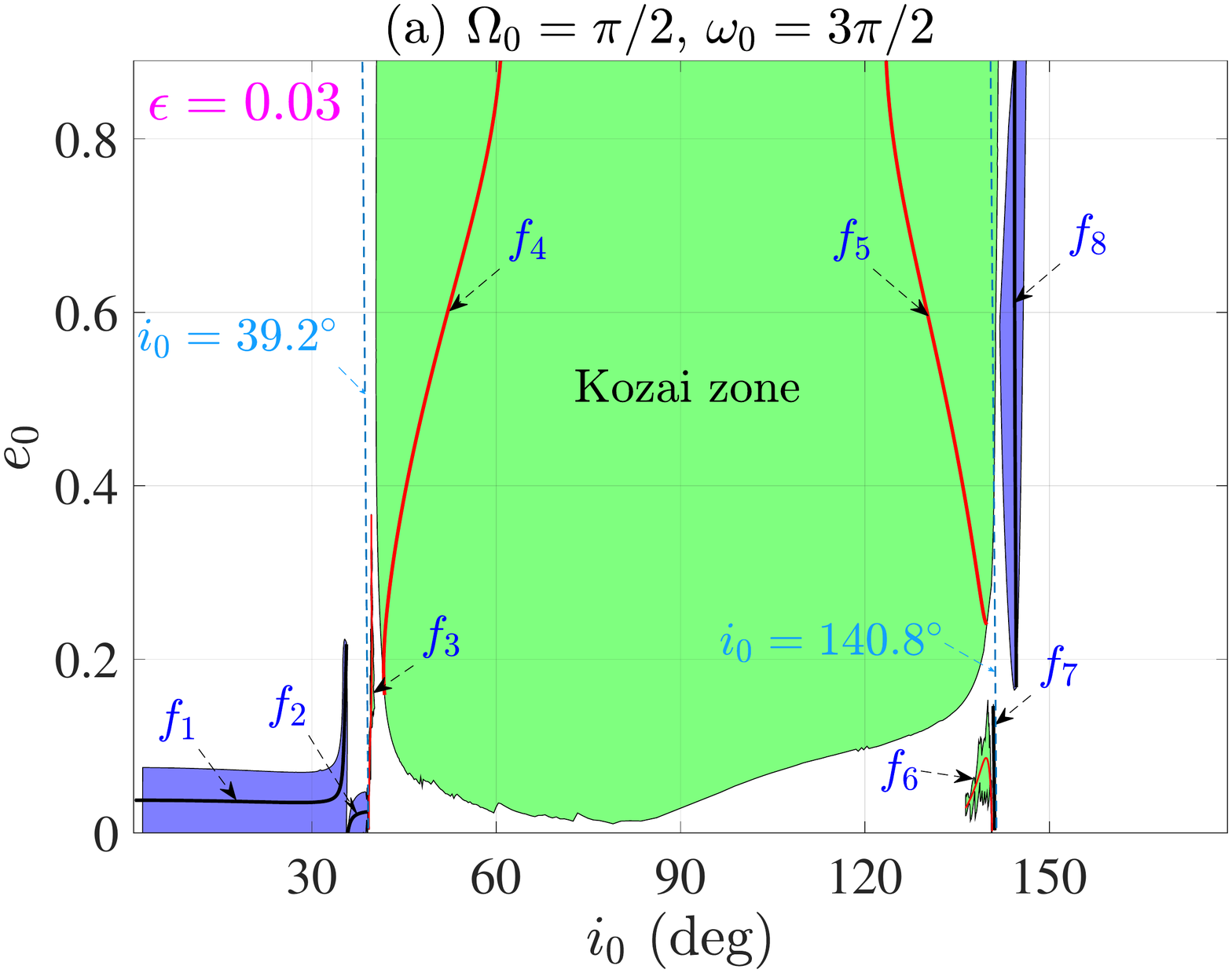}
\includegraphics[width=0.8\textwidth]{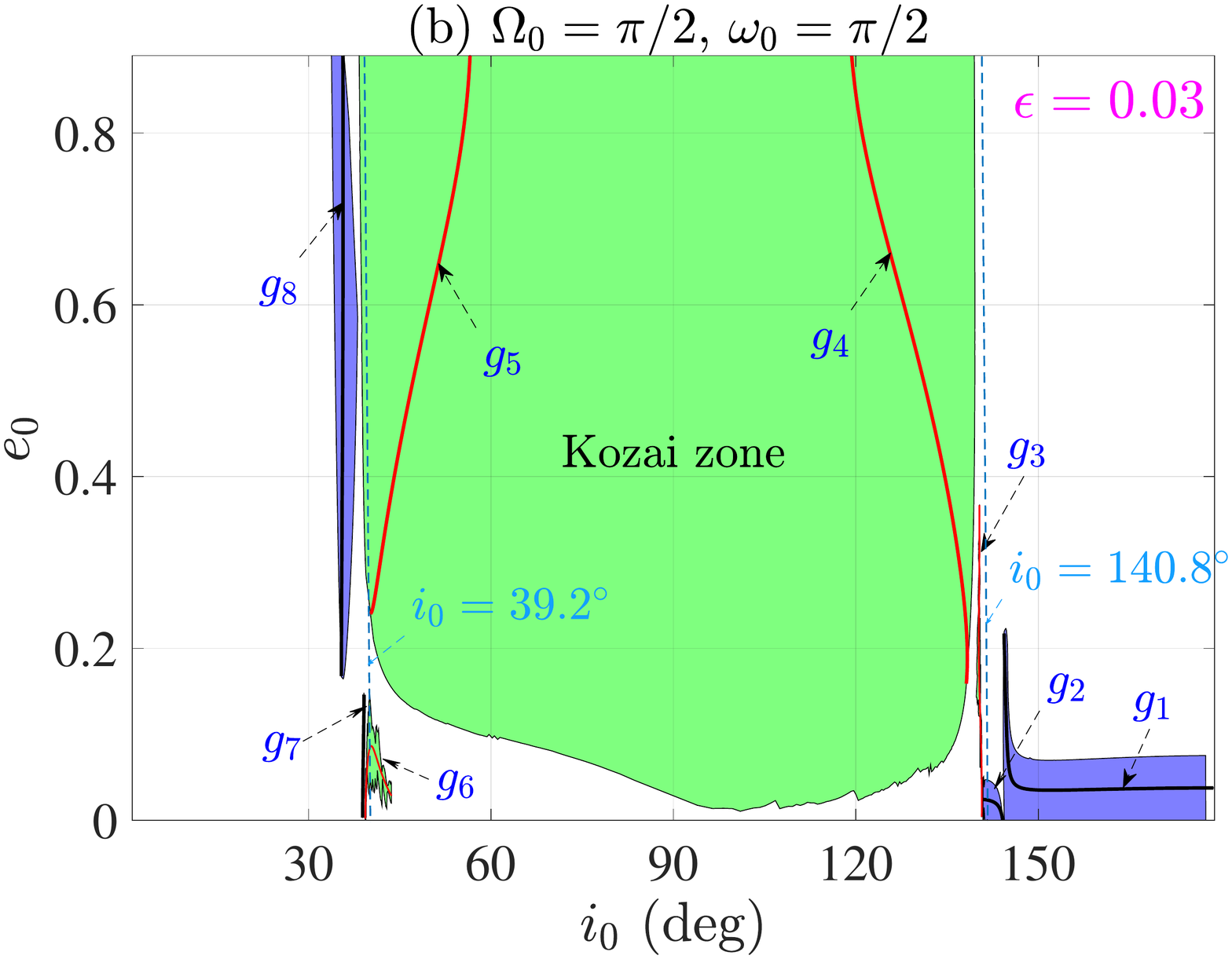}
\includegraphics[width=0.8\textwidth]{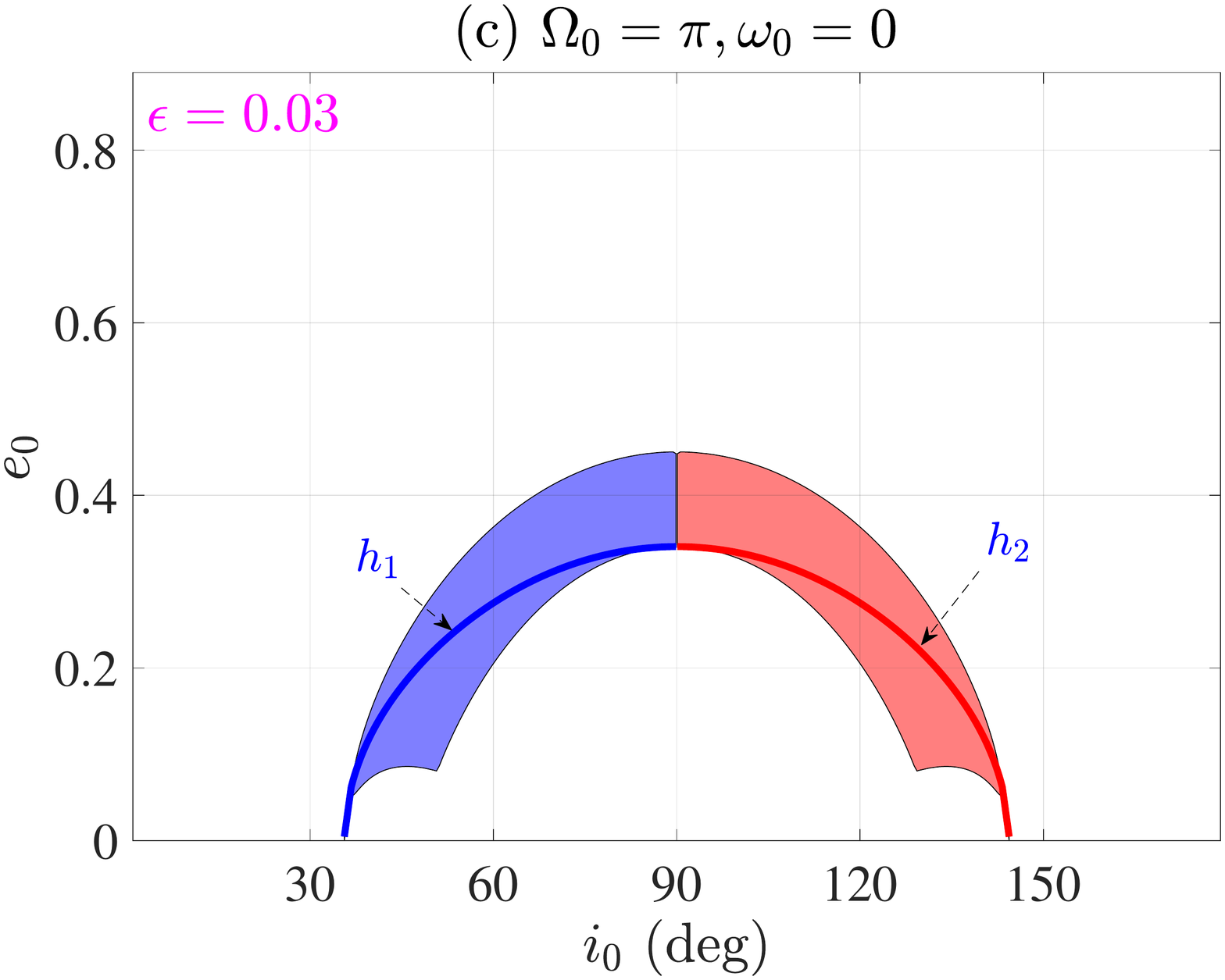}
\caption{Stable libration zones under dynamical model characterised by $\epsilon = 0.03$. The dashed lines specified by $i_0 = 39.2^{\circ}$ and $i_0 = 140.8^{\circ}$ stand for the separatrices of Kozai zones in the case of $\epsilon = 0$.}
\label{Fig5}
\end{figure*}

In the case of $\epsilon = 0.03$, the results are reported in Fig. \ref{Fig5}, where the stable libration zones are shown in the space of $(e, i)$ as shaded areas for three cases of symmetric periodic orbits with angles at $\Omega_0 = \pi/2$ and $\omega_0 = 3\pi/2$ (see panel a), $\Omega_0 = \pi/2$ and $\omega_0 = \pi/2$ (see panel b) and $\Omega_0 = \pi$ and $\omega_0 = 0$ (see panel c). The characteristic curves standing for the location of resonant centre are also shown in each panel. For convenience of comparison, the boundaries of Kozai zones in the case of $\epsilon = 0$ are shown in dashed lines, which are specified by $i_0 = 39.2^{\circ}$ in the prograde region and $i_0 = 140.8^{\circ}$ in the retrograde region.

From Fig. \ref{Fig5}, we can observe that (a) the secular resonances associated with $f_3 (g_3)$, $f_6 (g_6)$ and $f_7 (g_7)$ happen in the vicinities of $i_0 = 39.2^{\circ}$ and $i_0 = 140.8^{\circ}$ and they have very small domains in the considered space, (b) the secular resonances associated with $f_{1,2}$ ($g_{1,2}$) occur in the low-eccentricity region with inclinations smaller than $i_0 \approx 39^{\circ}$ (greater than $i_0 \approx 141^{\circ}$), (c) there is a large Kozai zone but it cannot cover the low-eccentricity region because the area near the zero-eccentricity point is filled with chaotic motion, as shown by the Poincar\'e sections in Figs. \ref{Fig1} and \ref{Fig2}, (d) in the high-eccentricity region the boundaries of Kozai zones move rightward (leftward) for case I (case II) in comparison to the ones in the case of $\epsilon =0$, (e) the resonances associated with $h_1$ and $h_2$ occurs in the mid- and low-eccentricity region (these two resonances do not exist in the high-eccentricity space), and (f) the libration zones associated with $h_1$ and $h_2$ are symmetric with respect to the polar line.

\begin{figure*}
\centering
\includegraphics[width=0.49\textwidth]{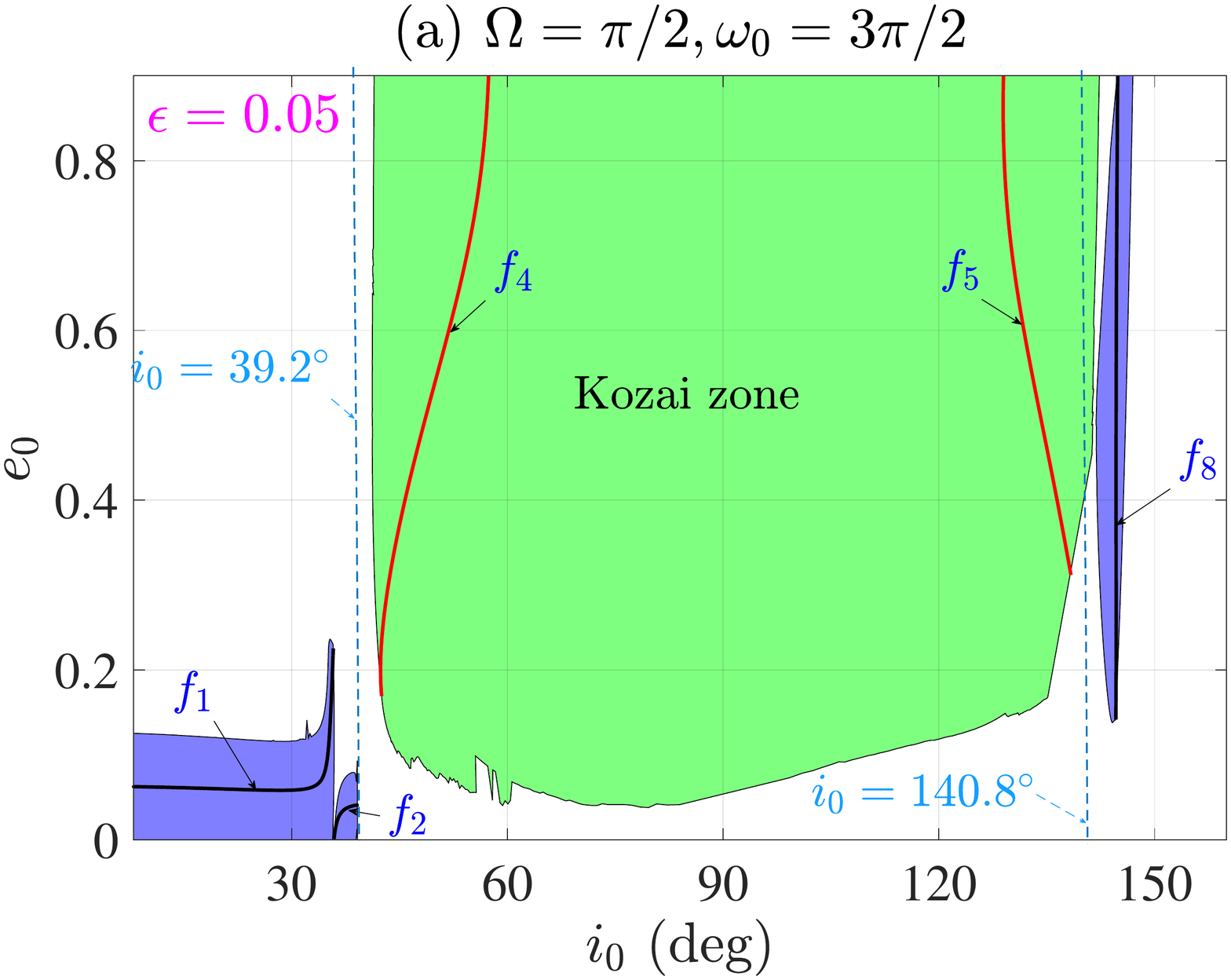}
\includegraphics[width=0.49\textwidth]{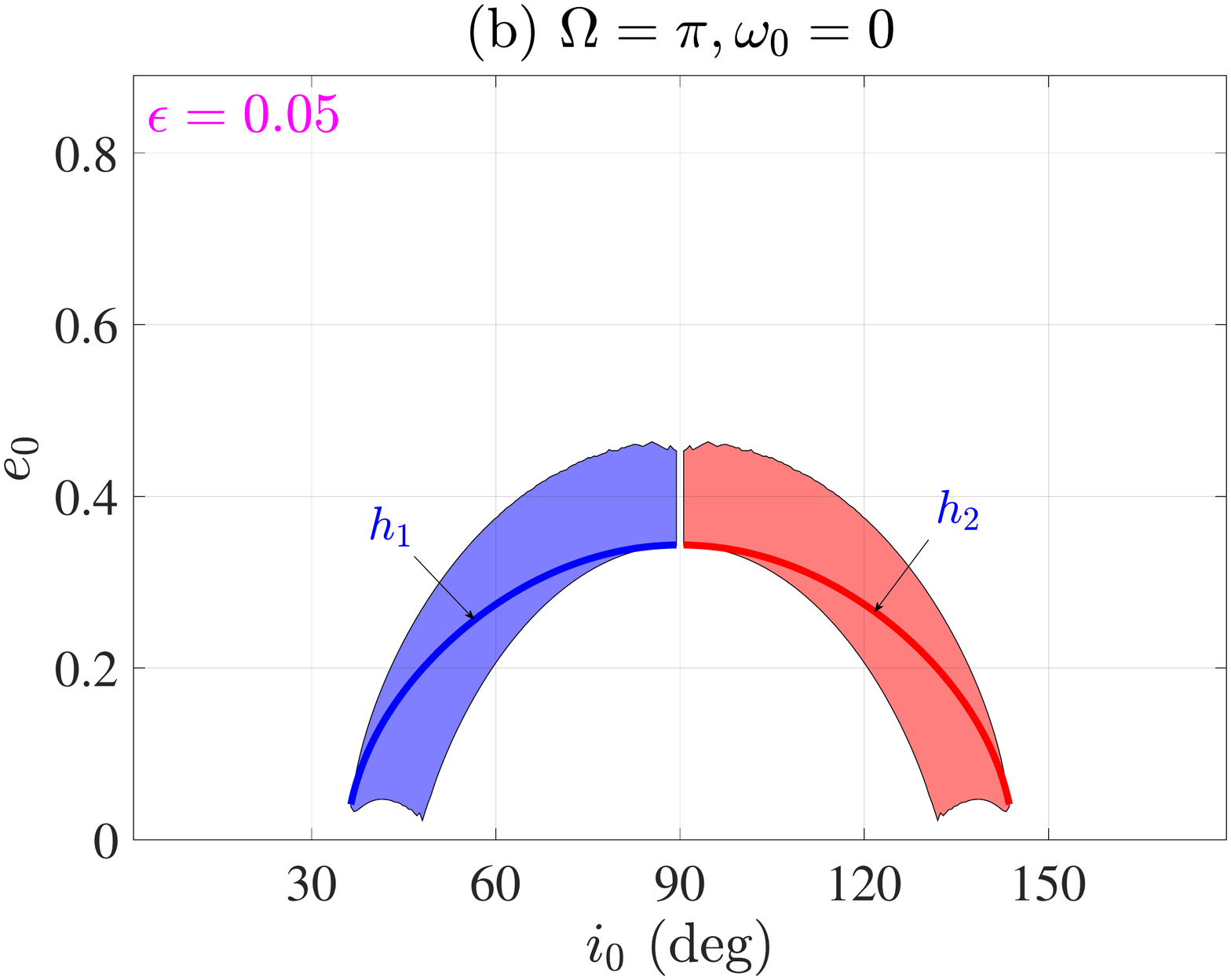}\\
\includegraphics[width=0.49\textwidth]{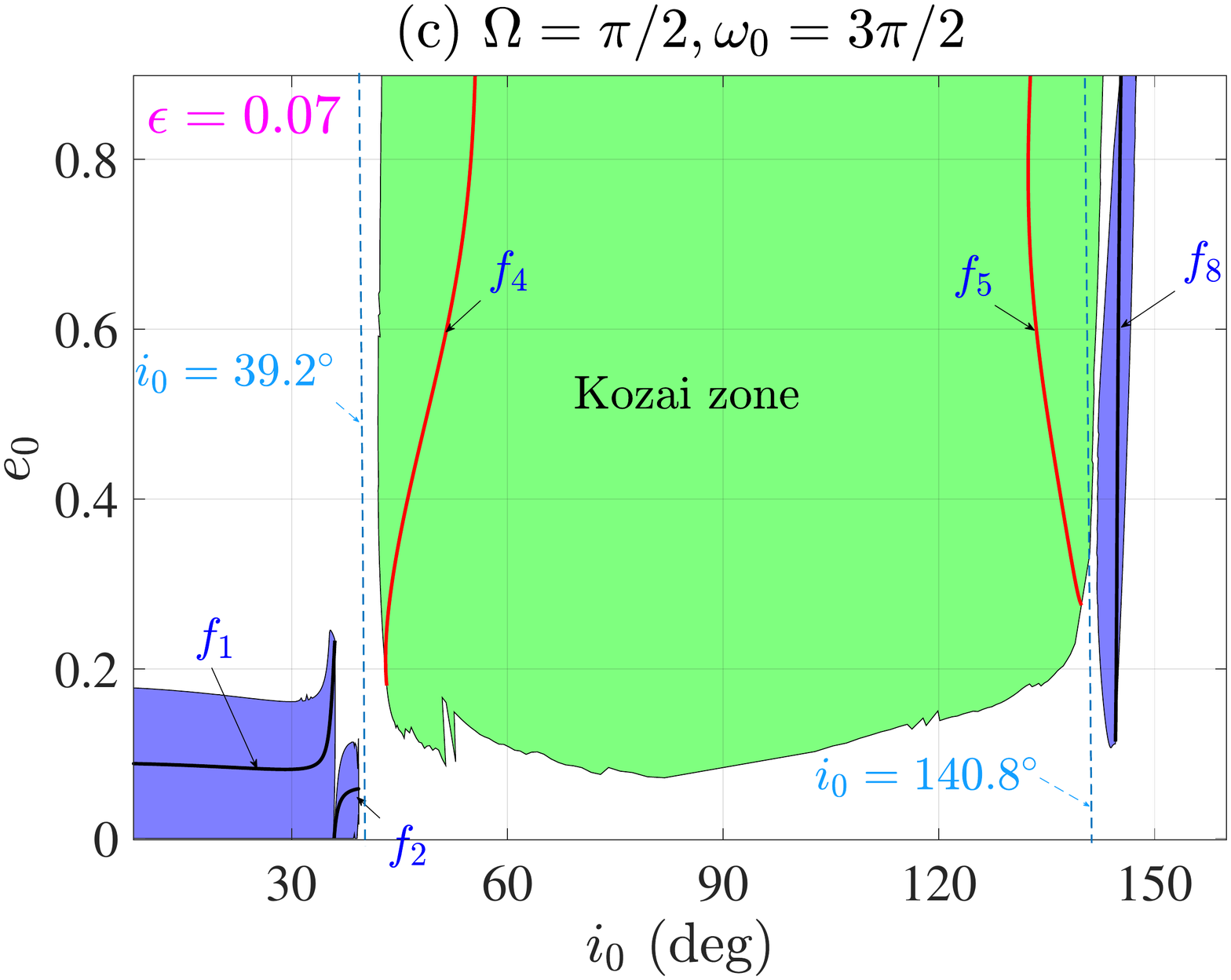}
\includegraphics[width=0.49\textwidth]{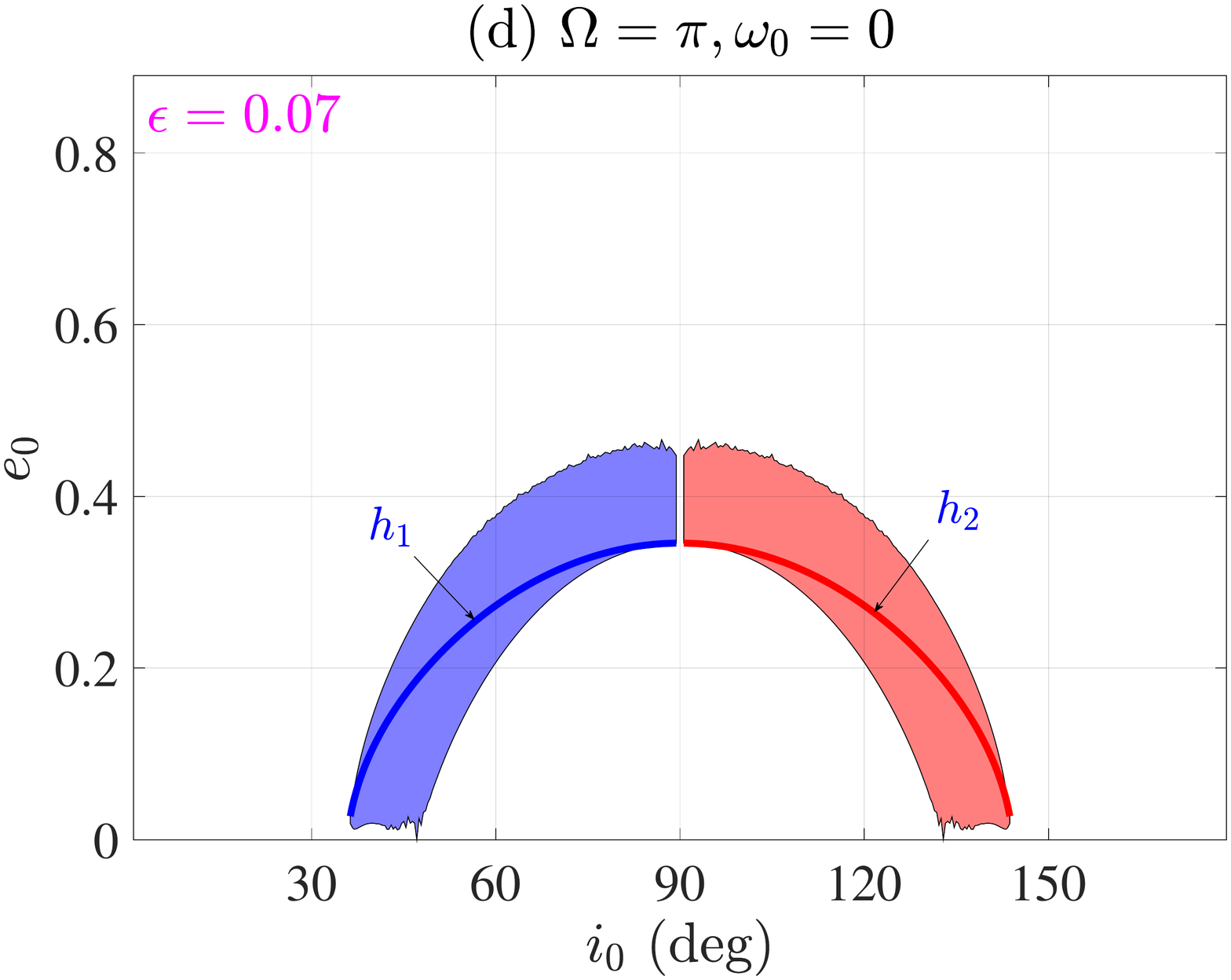}\\
\includegraphics[width=0.49\textwidth]{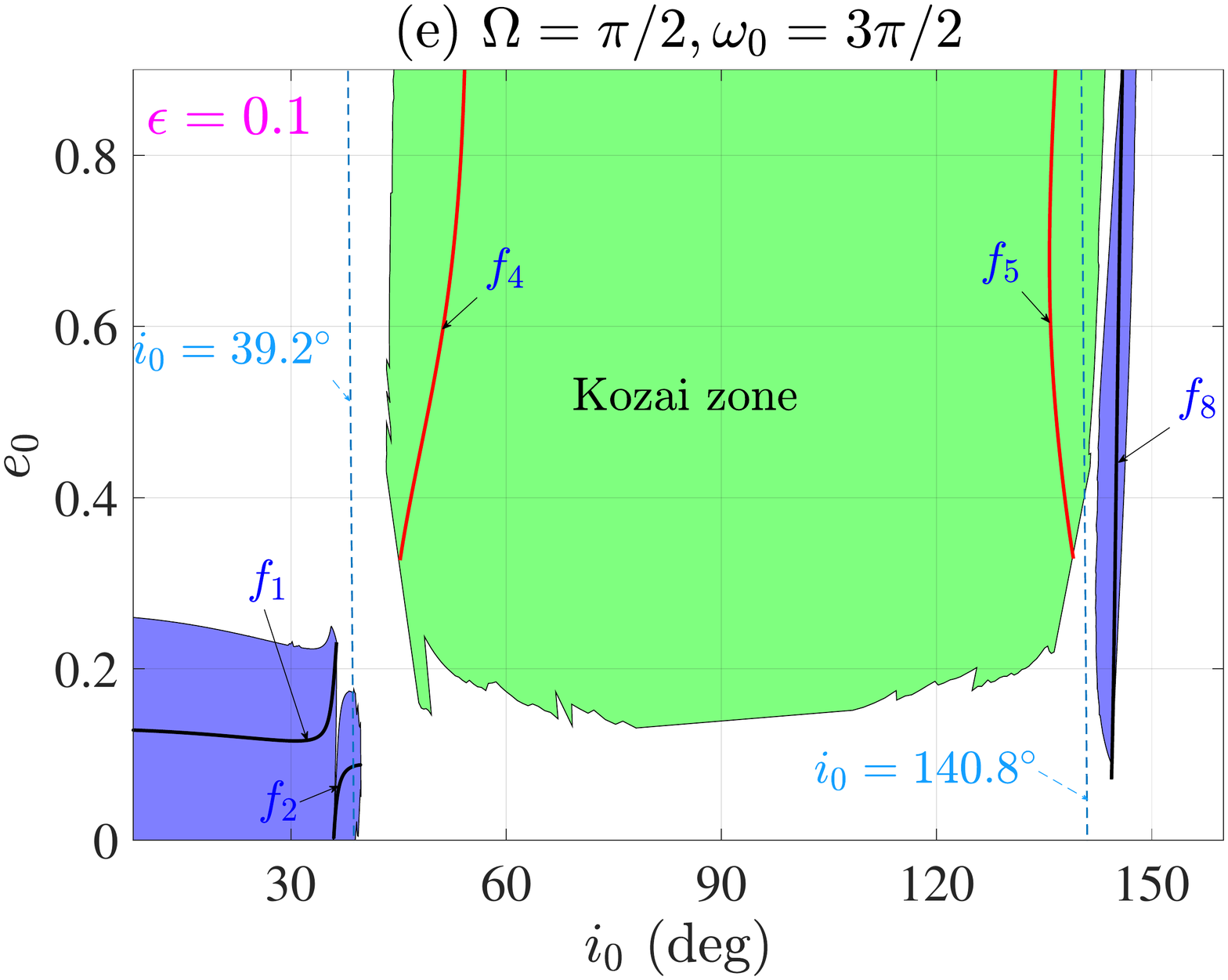}
\includegraphics[width=0.49\textwidth]{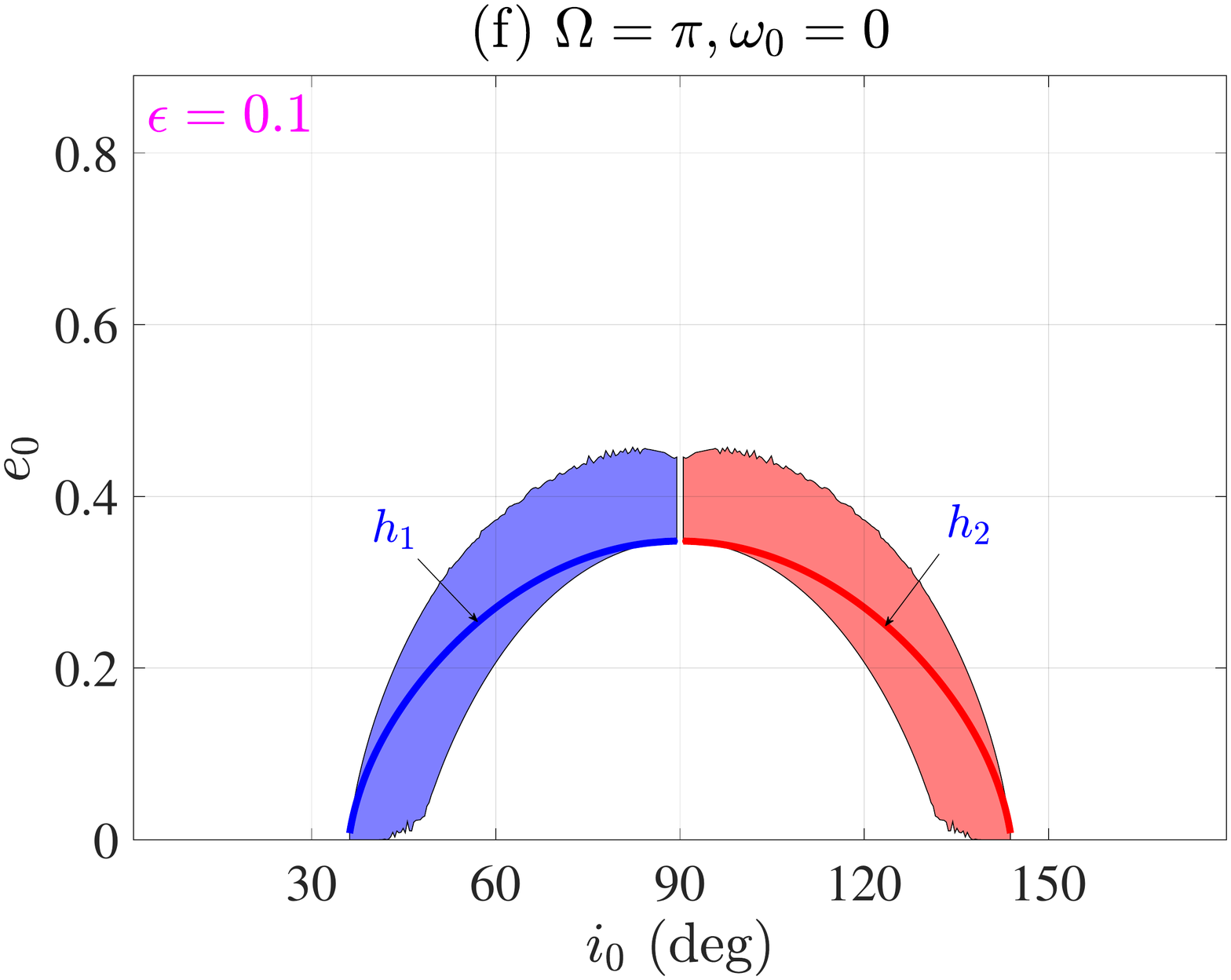}
\caption{Stable libration zones under dynamical models characterised by $\epsilon = 0.05, 0.07, 0.1$. The dashed lines specified by $i_0 = 39.2^{\circ}$ and $i_0 = 140.8^{\circ}$ stand for the separatrices of Kozai zones in the case of $\epsilon = 0$. The panels in the left column correspond to case I and the ones in the rigth column stands for case III (results about secular resonances in case II can be easily produced by means of symmetry).}
\label{Fig10}
\end{figure*}

Extending to higher values of $\epsilon$, we produce the boundaries of stable libration zones in the $(i_0, e_0)$ space, as shown in Fig. \ref{Fig10} where the dashed lines stand for the boundaries of Kozai zone in the case of $\epsilon = 0$. In practice, the cases of $\epsilon = 0.05, 0.07, 0.1$ are taken into account, and families $f_3$, $f_6$ and $f_7$ are not considered due to their small strength. From Fig. \ref{Fig10}, we can see that (a) the topological structures of libration zones for the primary resonances remain similar to the ones shown in Fig. \ref{Fig9}, (b) when $\epsilon$ increases, the area of Kozai libration zone becomes smaller and smaller, showing that the gap in low-eccentricity region becomes larger and larger, (c) libration zones associated with $f_1$ and $f_2$ have larger area with higher $\epsilon$, and (d) the libration zones associated with $f_8$, $h_1$ and $h_2$ are insensitive to the value of $\epsilon$.

Regarding the the apsidal secular resonances in families $f_1$ and $f_2$, \citet{lei2021} analytically investigated their dynamics using the perturbation theory based on Lie-series transformation and concluded that the considered resonances govern long-term evolutions of test particles with inclinations smaller than $39^{\circ}$. Comparing to the numerical results in the current work, we can see a good agreement between the stable libration zones associated with $f_1$ and $f_2$ (see Figs. \ref{Fig5} and \ref{Fig10}) and the analytical resonant width presented by \citet{lei2021} (see Fig. 6 in his work).

\section{Conclusions}
\label{Sect6}

In this work, we systematically investigated varieties of secular resonances for inner test particles in hierarchical planetary systems by means of non-perturbative approaches. To formulate the dynamical model for long-term evolution, the double-averaged Hamiltonian truncated at the third order in the semimajor axis ratio (the octupole-level approximation) is adopted. The resulting dynamical model is of two degrees of freedom with $\omega$ and $\Omega$ as the angular coordinates.

In order to obtain preliminary dynamics, we produced Poincar\'e surfaces of sections at different levels of Hamiltonian. In the surfaces of section, some structures are observed: some areas are filled with chaotic motion, some areas are filled with resonant trajectories (quasi-periodic orbits) and some areas are filled with circulatory trajectories. According to dynamical system theory, at the centre of each libration island (corresponding to the resonant centre), there is a stable periodic orbit. In general, we summarised three cases of symmetric periodic orbits as follows: case I with $\omega_0 = 3\pi/2$ and $\Omega_0 = \pi/2$, case II with $\omega_0 = \pi/2$ and $\Omega_0 = \pi/2$ and case III with $\omega_0 = 0$ and $\Omega_0 = \pi$.

Using numerical correction and continuation techniques, we computed families of stable and symmetric periodic orbits. The characteristic curves of families of stable periodic orbits are shown in the space $(e_0, i_0)$ and these curves are also named webs of secular resonances. To measure the boundaries of stable libration zones, we numerically determined the upper and lower limits in the action space by analysing Poincar\'e surfaces of section. The parameter spaces enclosed by the boundaries are called stable libration zones, inside which the trajectories are of libration during the period of numerical integration. The resonances considered in this study can be classified as aligned or anti-aligned apsidal secular resonances, Kozai resonances and high-order secular resonances. It should be mentioned that in this work only the resonances which are visible in the Poincar\'e surfaces of section are taken into account (those weak resonances are not considered). This is based on the consideration that they are the strongest from the viewpoint of strength. In addition, the boundaries of stable libration zones are obtained by analysing Poincar\'e sections, so results will be changed if different definitions of Poincar\'e section are adopted.

The effects of the octupole potential are evaluated by taking different values of $\epsilon$ ($\epsilon$ specifies the contribution of octupole terms). Under the quadrupole-order approximation (corresponding to $\epsilon = 0$), the dynamical model is totally integrable and there is only Kozai resonance, which occurs in the inclination space ranging from $39.2^{\circ}$ to $140.8^{\circ}$. Under the octupole-level approximation with non-zero $\epsilon$, besides (perturbed) Kozai resonance, new secular resonances (such as apsidal resonances and high-order resonances) appear in the phase space and all these resonances constitute a web (called web of resonances). Due to the perturbation of octupole-order potential, dynamics of Kozai resonance is changed: Kozai resonance disappears in the low-eccentricity region because this zone is filled with chaos. Results show that the chaotic area increases with the factor $\epsilon$.

The results obtained in this work are helpful to understand the secular dynamics of hierarchical exoplanetary systems in terms of the following concerns: (a) how many kinds of secular resonances existing in the action space, (b) the place where the exact resonance occurs, (c) the region where a certain secular resonance can occur or not, and (d) the size of libration zone changing with eccentricity and/or inclination for a particular resonance.

\begin{acknowledgements}
The author thanks two anonymous reviewers for their insightful and rigorous comments that help to improve the quality of this paper. This work is performed with the financial support of the National Natural Science Foundation of China (No. 12073011).
\end{acknowledgements}


\section*{Conflict of interest}
The authors declare that they have no conflict of interest.

\bibliographystyle{spbasic}      
\bibliography{mybib}   

%
%

\end{document}